\documentclass[a4paper,12pt]{article}
\usepackage[english]{babel}
\usepackage{url}
\usepackage[utf8x]{inputenc}
\usepackage{amsmath}
\usepackage{graphicx}
\graphicspath{{images/}}
\usepackage{enumitem}
\usepackage{siunitx}
\usepackage{hepnames}
\usepackage{fancyhdr}
\usepackage{booktabs}
\usepackage{vmargin}
\usepackage{microtype}
\usepackage{graphicx}	
\usepackage[colorinlistoftodos]{todonotes}
\usepackage{hepnames}
\usepackage{braket}
\usepackage{epstopdf}
\usepackage[version=4]{mhchem}
\usepackage{pifont}
\usepackage{xcolor}
\usepackage{caption}
\usepackage{subcaption}
\usepackage{braket}
\usepackage{nth}
\usepackage[version=4]{mhchem}
\usepackage{geometry}
\usepackage[pdftex,
            pdfauthor={William G. S. Vinning},
            pdftitle={Combining Electron Appearance and Muon Disappearance Channels in Light Sterile Neutrino Oscillation Searches at Fermilab's Short-Baseline Facility}]{hyperref}
\hypersetup{
    colorlinks,
    linkcolor={black},
    citecolor={black},
    urlcolor={blue!80!black}
}

\DeclareSIUnit\c{c}
\DeclareSIUnit\ppb{ppb}
\DeclareSIUnit\POT{POT}
\DeclareSIUnit\foot{ft}
\newcommand{\PRLsep}{\noindent\makebox[\linewidth]{\resizebox{0.3333\linewidth}{1pt}{$\bullet$}}\bigskip}

\newcommand\tab[1][1cm]{\hspace*{#1}}
\numberwithin{equation}{section}

\setmarginsrb{3 cm}{2.5 cm}{3 cm}{2.5 cm}{1 cm}{1.5 cm}{1 cm}{1.5 cm}

\title{Student No: 33379165}								
\author{William Vinning}								
\date{\today}											

\makeatletter
\let\thetitle\@title
\let\theauthor\@author
\let\thedate\@date
\makeatother

\begin{document}

\pagenumbering{roman}
\begin{titlepage}
\thispagestyle{empty}
\begin{center}
\rule[0.5ex]{\linewidth}{2pt}\vspace*{-\baselineskip}\vspace*{3.2pt}
\rule[0.5ex]{\linewidth}{1pt}\\[-0.1 cm]
{\huge \bfseries Combining \Pnue Appearance and \Pnum Disappearance Channels in Light\\[-0.00cm] Sterile Neutrino Oscillation \\[-0.02cm] Searches at Fermilab's \\[0.36cm]Short-Baseline Neutrino Facility}\\[0.1cm]
\rule[0.5ex]{\linewidth}{1pt}\vspace*{-\baselineskip}\vspace*{4.2pt}
\rule[0.5ex]{\linewidth}{2pt}\\
\vspace{6.5mm}
{\large By}\\
\vspace{6.5mm}
{\large\textsc{William G. S. Vinning}}\\
\vspace{11mm}
{\large Supervised by}\\
\vspace{6.5mm}
{\large\textsc{Dr. Andrew Blake}}\\
\end{center}

	\centering
    \vspace*{0.5 cm}
    \includegraphics[scale = 0.16]{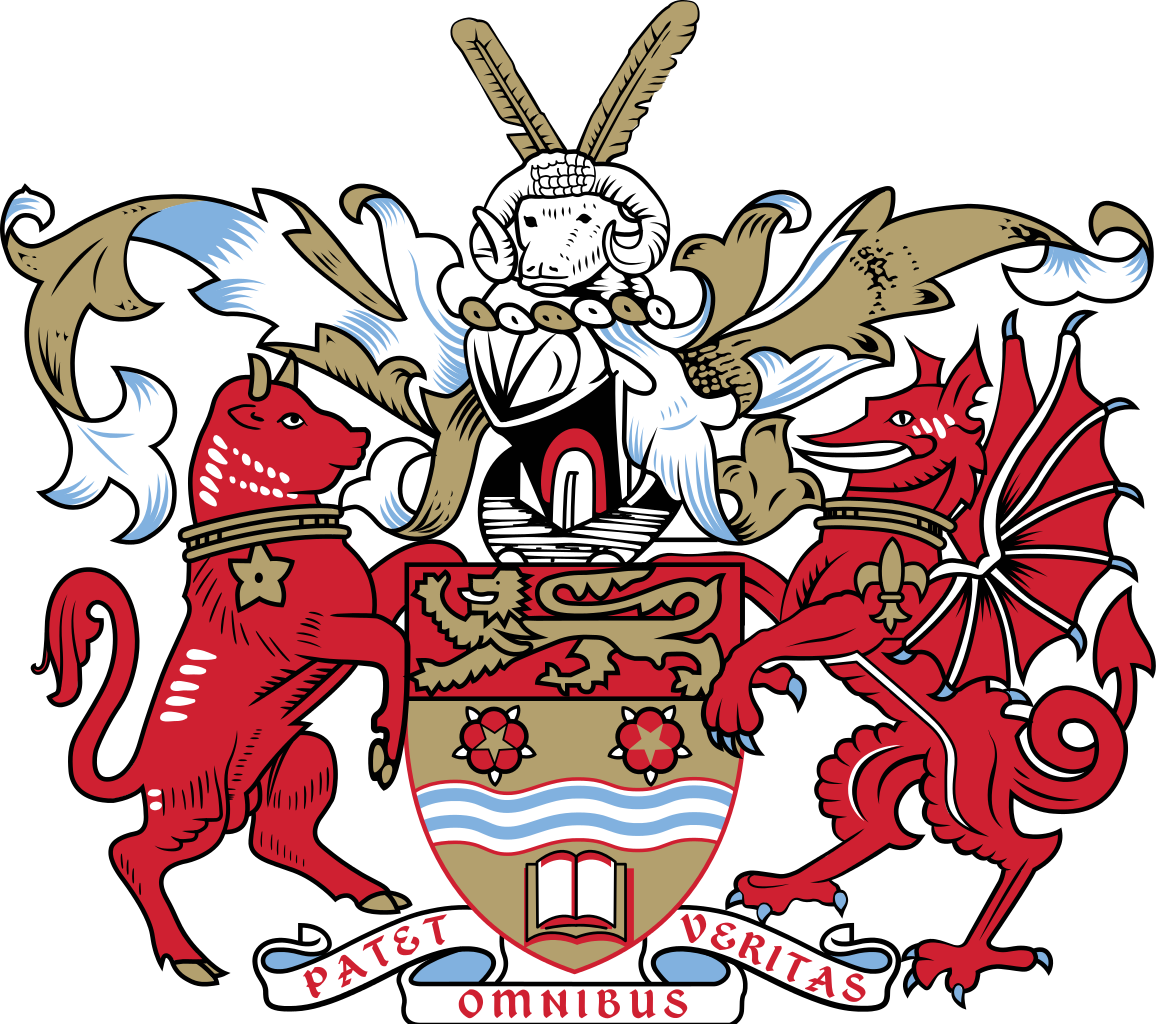}\\[0.5 cm]	
		 \large Department of Physics\\[0.2 cm]
    \textsc{\Large Lancaster University}\\
			\vspace*{\fill}
	\textsc{\Huge}				
	
	\newpage
	\thispagestyle{empty}
\begin{abstract}
    Of late, a number of instances of neutrino flux anomalies observed at short-baselines have given traction to the hypothesis of adding new neutrino flavours to our Standard Model set, albeit ones not associated with a partnered lepton. Anomalies observed at LSND, MiniBooNE and at short distances from nuclear reactors have suggested the existence of sterile mass states with masses on the scale of $\sim\SI{1}{\electronvolt}$, prompting further investigations. Subsequently, Fermilab is set to host a Short-Baseline Neutrino (SBN) oscillation program which will cross-check each of these anomalies by deploying three state-of-the-art liquid argon detectors along the Booster Neutrino Beamline. Through an event-by-event Monte Carlo simulation study, this document presents short-baseline oscillation sensitivity predictions for SBN on a purely statistical basis in the scope of a 3+1 sterile neutrino model approximated with a two flavour mixing basis. As a world's first, sensitivities in parameter space to 3+1 mixing angles $\theta_{14}$ and $\theta_{24}$ are presented, in addition to the sensitivities gained by combining \Pnum disappearance and \Pnue appearance channel data. Provided also are energy spectra and 3+1 oscillation parameter sensitivity predictions for these two individual channels. The results agreed with parameter space sensitivity predictions already conceived by Fermilab to a satisfactory degree.
\end{abstract}

\end{titlepage}
\newpage
\thispagestyle{empty}
\setcounter{page}{3}
\renewcommand{\abstractname}{Acknowledgements}
\begin{abstract}
I would like to extend the greatest thanks to my supervisor, Andy, for always providing a generous and invaluable insight whenever I have needed. I am grateful also to Dr. Dom Brailsford for passing on his decay point distribution analysis. Thanks also to Dr. Bill Louis of Los Alamos National Laboratory for passing on the unicorn-like LSND data. I would like to extend a special thanks to my high school physics teacher, Jim Barry, for subliminally planting my interest in neutrino oscillations with his influential `Ali-G' wave.
\end{abstract}
\newpage
\pagestyle{empty}
\listoffigures
\thispagestyle{empty}
\thispagestyle{empty}
\newpage
\thispagestyle{empty}
\newpage
\thispagestyle{empty}
\tableofcontents
\thispagestyle{empty}

\pagebreak


\pagestyle{fancy}
\fancyhf{}
\fancyhead[RE,LO]{\thepage}
\fancyhead[LE,RO]{\rightmark}
\newgeometry{twoside,a4paper}
\setmarginsrb{3 cm}{2.5 cm}{3 cm}{2.5 cm}{1 cm}{1.5 cm}{1 cm}{1.5 cm}
\section{Introduction}
\pagenumbering{arabic}

The lepton sector in the frame of the Standard Model (SM) embodies the outlook of 3 neutrino flavours, $\Pnue$, $\Pnum$ and $\Pnut$; which are assumed to be massless from the Dirac symmetry breaking approach. Neutrinos are however confirmed to harbour mass, albeit very small, with combined magnitude estimated to be equal to less than one millionth of that of an electron from cosmological bounds\cite{mass}. Hence their name, neutrinos are neutral particles and interact with matter exclusively via the weak interaction, a process mediated by the W and Z bosons. This fact, combined with their tiny masses, renders neutrinos to be incredibly elusive to even modern detectors as they intrinsically interact very rarely with matter, with interaction cross-sections on the scale of $\sim$ \SI{e-38}{\cm\squared}.

The notion of massive neutrinos is also inferred from the observation of flavour oscillations, in which a neutrino of a particular flavour bears the potential to transition to another flavour after traversing some distance. The confirmation of this non-standard phenomenon was the subject of the Nobel Prize for Physics in 2015 regarding projects at Super-Kamiokande\cite{superk} and SNO\cite{sno1}. Flavour transitions, and subsequently lepton number violation, were first speculated in 1957 by Brian Portecorvo as a characteristic property of introducing mass into the neutrino sector\cite{portecorvo,portecorvo1}. We now identify the interacting SM neutrinos, with well-defined flavours, to be `weak' eigenstates and rather a construction of a quantum mechanical superposition of `mass' eigenstates (and vice versa), namely $\nu_1$, $\nu_2$ and $\nu_3$, which possess well-defined masses. In other words, neutrino flavours are allowed to mix in a similar fashion to that of quarks at strengths represented by the relevant elements of the \num{3 x 3} unitary PMNS mixing matrix, expressed as $U$ in Equation \ref{eq:rep}.

	\begin{align}\label{eq:rep}
		\ket{\nu_{\alpha}} &= U \ket{\nu_i} & \alpha &= e, \mu, \tau \tab i = 1, 2, 3
	\end{align}
	The PMNS matrix is often deconstructed into a representation parametrised by a set of explicit phases, or mixing angles, as shown in Equation \ref{eq:PMNS}.
\begin{center}
	\begin{equation} \label{eq:PMNS}
	U = 		\begin{pmatrix}
		 c_{12}c_{13}& s_{12}c_{13} & s_{13}e^{-i\delta_{CP}}\\ 
		 -s_{12}c_{23}-c_{12}s_{23}s_{13}e^{i\delta_{CP}}& c_{12}c_{23}-s_{12}s_{23}s_{13}e^{i\delta_{CP}} & s_{23}c_{13}\\ 
		 s_{12}s_{23}-c_{12}c_{23}s_{13}e^{i\delta_{CP}}& -c_{12}s_{23}-s_{12}c_{23}s_{13}e^{i\delta_{CP}}  & c_{23}c_{13}
		\end{pmatrix}
	\begin{pmatrix}
	1 & 0 & 0 \\
	0 & e^{i\frac{\alpha_{1}}{2}} & 0 \\
	0 & 0 & e^{i\frac{\alpha_{2}}{2}}
	\end{pmatrix}
	\end{equation}
\end{center}
where $s_{ij} \equiv \sin\theta_{ij}$ and $c_{ij} \equiv \cos\theta_{ij}$, in which $\theta_{ij}$ is the relative mixing angle between mass states $i$ and $j$. Additionally, $\delta_{CP}$ represents a charge-parity (CP) violating phase and $\alpha_{i}$ is a Majorana phase, both currently undetermined.

    Neutrino oscillations arise from considering the propagation of neutrino mass eigenstates as plane wave solutions to the time-dependent Schrödinger equation, in which their phase velocity is proportional to their mass value. Flavour states are ultimately a superposition of mass eigenstates with non-degenerate mass values, therefore as a neutrino propagates through space, interference occurs and the relative admixture of mass eigenstates changes. This mechanism results in a possible flavour transition, bearing in mind the probabilistic nature of quantum mechanics. The probability of transition from flavour states $\Pnu_{\alpha}$ to $\Pnu_{\beta}$ (where $\alpha \neq \beta$), in a simplified two flavour model is expressed by Equation \ref{eq:osc}.
	
	\begin{equation}
		\label{eq:osc}
	P_{\nu_\alpha \rightarrow \nu_\beta} = \sin^2\left(2\theta\right)\sin^2 \left( \frac{\Delta m^2L}{4E}\right)
	\end{equation}

The transition probability proceeds as an oscillatory function of the traversed distance $L$, at a rate decided by the ratio of the difference in mass squared $\Delta m^2$ between the two complementary mass states to the neutrino energy $E$. Moreover, the amplitude of oscillation is set by a function of the mixing angle $\theta$ between mass states $\nu_1$ and $\nu_2$.

Obtaining the oscillation parameter values to a satisfactory precision has been a prime focus of the neutrino sector since oscillations were confirmed, of which the current best measurements are detailed in Table \ref{tab:param}.

\begin{table}[h!]
\caption{Current best measurements of the standard oscillation parameters\cite{pdg}.} \label{tab:param}
    \centering
    	\begin{tabular}{ccl} \hline
    	Parameter & Value & ~ \\ \hline
    	\\[-1em]
    	$\Delta m^2_{21}$ & \SI[separate-uncertainty=true]{7.53\pm0.18 e-5}{\electronvolt\squared}& ~\\
   	$\left|\Delta m^2_{32}\right|$ & \SI[separate-uncertainty=true]{2.44\pm0.06 e-3}{\electronvolt\squared}& (normal MH) \\
    	 & \SI[separate-uncertainty=true]{2.51\pm0.06 e-3}{\electronvolt\squared} & (inverted MH) \\
    	$\sin^2(\theta_{12})$ & $0.304 \pm 0.014$ & ~\\
    	$\sin^2(\theta_{23})$ & $0.51 \pm 0.05$ & (normal MH) \\
    	 & $0.50 \pm 0.05$ & (inverted MH) \\
    	$\sin^2(\theta_{13})$ & $(2.19 \pm 0.12) \times 10^{-2}$ & ~ \\ \hline
    	\end{tabular}
\end{table}

Though being a relatively new field, much of the physics describing neutrinos is still mysterious. For instance, the mass hierarchy (MH), or the relative mass ordering of mass states, is yet to be determined as the sign of $\Delta m^2_{32}$ is unknown. The next generation of very long-baseline oscillation experiments, including DUNE\cite{dune} and NO$\nu$A\cite{nova}, is set to provide the first measurements of this. The absolute masses of neutrino flavours are also not known, though cosmological limits to the combined mass are set. CP violation in the neutrino sector seems to be favoured by observations\cite{t2k}, though precision measurements of $\delta_{CP}$ are not yet available. Efforts at the long-baseline $\theta_{13}$ sensitive experiment T2K are shifting to prioritise CP phase measurement, with greater prospects coming with the approaching Hyper-Kamiokande upgrade\cite{hyperk}. As well, though inconsequential to oscillation physics, neutrinos could possibly be Majorana particles, rendering neutrinos to be their own anti-particles. The observation of a non-zero rate of the radioactive decay process, neutrinoless double beta decay\cite{n0bb}, will be indicative of a Majorana scenario, of which there are searches at SNO+\cite{sno} and SuperNEMO\cite{supernemo}. Perhaps most of all, there is no known mechanism for the acquisition of mass in the scope of a 3$\nu$ model where neutrinos and anti-neutrinos are distinct.

Moreover, in recent years, several experimental instances have suggested the presence of oscillations occurring at much lower values of $L/E$ ($\sim$\SI{1}{\per\electronvolt\squared}), inconsistent with oscillation scales set by the standard mixing parameters. These Short-Baseline (SBL) experimental anomalies can be summarised as such:

\begin{itemize}[noitemsep]
\item In SBL accelerator experiments, an observation of an excess of \Pnue appearing in a predominantly \Pnum beam in anti-neutrino mode at LSND to $3.8\sigma$\cite{lsnd} and both neutrino and anti-neutrino modes at MiniBooNE to $3.4\sigma$ and $2.8\sigma$\cite{miniboone} respectively.
\item In various SBL nuclear reactor experiments, an observed deficit of \APnue with a combined observed to predicted flux ratio of $0.943 \pm 0.023$, a deviation from unity at $2.5\sigma$\cite{reactoranomaly}.
\item In using gallium to calibrate solar neutrino detectors, a deficit of \APnue observed at experiments GALLEX\cite{gallex} and SAGE\cite{sage} at short distances with a combined statistical significance of $3.0\sigma$\cite{gallium}.
\end{itemize}

These observations could be interpreted as new oscillation modes generated by a new neutrino of the eV mass scale, though with properties dissimilar to that of the standard `active' flavours which interact weakly. Inspections of the Z boson decay width from $\Ppositron \Pelectron$ collider experiments place the number of active flavours at $2.9840 \pm 0.0082$\cite{pdg}, hence there is motivation surrounding an extension of the SM with neutrinos which are exempt from this process. These `sterile' neutrino flavours are theorised to interact only via gravity and the Higgs mechanism so their presence may be inferred only by their generation of non-standard oscillation signatures.

Conversely, there are various sterile searches which were sufficiently consistent with a standard 3$\nu$ mixing model.

\begin{itemize}[noitemsep]
\item In SBL accelerator experiments, no observation of anomalous $\Pnum$ disappearance at SciBooNE\cite{sciboone} and \APnue appearance in KARMEN\cite{karmen}.
\item In medium-baseline $\theta_{13}$ measuring reactor experiment, Daya Bay\cite{dayabay}, no observation of an anomalous \APnue deficit.
\item In long-baseline experiments IceCube\cite{icecube} and MINOS\cite{minos}, no measurement of an anomalous deficit of both \Pnum or \APnum.
\item In the Planck 2015 cosmological review\cite{planck}, a measurement of the number of relativistic degrees of freedom of $N_{\text{eff}} = 3.15\pm0.23$, favouring a 3$\nu$ model.
\end{itemize}

Nevertheless, providing clarity to these SBL anomalies is a significant point of interest at current. A Short-Baseline Neutrino (SBN) program hosted at Fermilab has been approved to provide new limits on sterile neutrino mixing and provide decisive refutations of the MiniBooNE and LSND excesses using state-of-the-art liquid argon detector technology. Through three detectors deployed along the Booster Neutrino beam, SBN will cross-check both the accelerator anomalies, by searching for both \Pnum disappearance and \Pnue appearance oscillation signatures, and the reactor anomalies, by searching for \Pnue disappearance. This report produces new sensitivity limits of SBN to SBL neutrino oscillations on a purely statistical basis through an event-by-event Monte Carlo simulation. An effort has been made to combine \Pnue appearance and \Pnum disappearance channels in order to provide a new sensitivity representation surrounding the $\theta_{14}$ and $\theta_{24}$ mixing parameters introduced by a 3+1 sterile neutrino model.

The structure of this report is as follows. Section 2 summarises theoretical perspectives of extending the SM with sterile neutrinos and provides more experimental detail on the observed SBL anomalies. Section 3 outlines the SBN program in technical details in addition to a rundown of basic LAr-TPC operation. An up-to-date summary of SBN's current status is detailed as well as this report's placement within the present landscape. Section 4 describes the preliminary information and procedure of the simulation, in terms of event generation and encoding of oscillations. Section 5 and 6 display and analyse the information available from \Pnum disappearance and \Pnue appearance studies respectively. Section 7 presents the full sensitivities to the 3+1 mixing angles of the SBN program from combining \Pnum disappearance and \Pnue appearance channels. In Section 8, conclusions are reached and further work is postulated. Appendix A shows the impact on energy spectra of the \Pnue disappearance channel and Appendix B shows the reduction in sensitivities from considering a distribution of oscillation baselines as opposed to fixed baselines.
\newpage
\section{Background}

\subsection{Sterile Extensions to the Standard Model}

As spin-1/2 particles, the dynamics of free neutrinos are described by the Dirac equation, with the Lagrangian density for the Dirac field as shown in Equation \ref{eq:diraceq}.
\begin{equation}\label{eq:diraceq}
\mathcal{L} = \overline{\psi}(i\gamma^\mu\partial_\mu - m)\psi
\end{equation}
where $\psi$ represents a neutrino field, $\gamma^\mu$ is the $\mu$th contravariant gamma matrix and $m$ is mass.

Expressing the neutrino field as a chiral spinor of Left-Handed (LH) and Right-Handed (RH) fields as in Equation \ref{eq:fuck}, the mass term of the neutrino Lagrangian may only be recovered by introducing a non-zero RH neutrino field, for which there exists two possible approaches\cite{rh}.

\begin{equation}\label{eq:fuck}
\psi = \begin{pmatrix} \psi_{L} \\ \psi_{R} \end{pmatrix}, \tab \overline{\psi} = \psi^{\dagger}\gamma^0
\end{equation}

A non-zero mass term may be generated solely using terms defined by the LH neutrino fields from the Majorana mechanism\cite{steveyboyd}. By requiring that the RH neutrino field is defined by $\psi_R = \hat{C}\overline{\psi_L}^\dagger \equiv \psi^c_L$, whereby $\hat{C}$ is the charge conjugation operator, it follows then that $\hat{C}\psi \equiv \psi^c = \psi$ and the mass term is of a form as described in Equation \ref{eq:majorana}.

\begin{equation}\label{eq:majorana}
\mathcal{L}_{Majorana} = -\frac{1}{2}m\left(\overline{\psi^c_L}\psi_L + \overline{\psi_L}\psi^c_L\right)
\end{equation}
This formalism can be introduced without adding any new degrees of freedom to the SM, though needs to be generated by a spontaneously symmetry breaking gauge invariant term to retain gauge invariance\cite{rh}.

Additionally, a mass term could be retained simply by the existence of RH neutrino fields, more commonly referred to as sterile neutrinos, as represented by Equation \ref{eq:dirac}.

\begin{equation}\label{eq:dirac}
\mathcal{L}_{Dirac} = -m\left(\overline{\psi_R} \psi_{L} + \overline{\psi_L}\psi_{R}\right) 
\end{equation}

The neutrino remains to be the only fermion with just a LH chirality experimentally confirmed, hence either RH neutrinos cease to exist in nature or they interact too weakly with matter to be observed. This second postulate aligns with the formalism of V-A theory, which enforces the gauge bosons of the weak interaction may only couple to LH fields, such that sterile neutrinos transform as gauge singlets under arbitrary electroweak SU(2) rotations. Nevertheless, RH neutrino fields could still participate in Yukawa interactions with the LH fields and the Higgs field, or maybe proceed by more exotic interactions. Sterile neutrinos are necessary to explain the vast differences in mass scales for neutrinos and their partnered complementary leptons under various see-saw mechanism models\cite{seesaw}. Seemingly, the most natural sterile extension to the SM is three RH neutrinos to reflect our three known LH flavours, though it could also be assumed there is only one which takes the role of a gauge particle and only exists to generate neutrino mass. A complete Left-Right symmetric lepton sector, which could be nested in a SO(10) grand unified theory but also low scales too, enables this mechanism by supplying SU(2)$_L$ and SU(2)$_R$ triplets with hypercharge Y = 2\cite{gut}. Undoubtedly, the range of theory surrounding RH neutrinos is rich and leaves many opportunities for new and interesting physics.

A `3+N' extension to the SM adds N sterile flavour and mass states, increasing the size of the PMNS matrix by N columns and rows. A sterile neutrino is postulated to possess low mixing with the standard mass states, hence being dominantly composed of the sterile mass state admixture with respect to it's flavour basis (and vice versa). Though it is impossible to directly detect stable sterile neutrinos, it's mixing can still be studied from oscillation analyses of the standard active flavours in the regular way but at $L/E$ scales dictated by the mass of the \nth{4} mass eigenstate in comparison to the standard neutrino masses. However, the horizontal sterile elements of the PMNS matrix ($U_{s1}, U_{s2}...U_{sn}$) may only be found through the assumption of unitary.


	
	
RH neutrinos of varying mass scales are a significant point of interest in the scope of cosmology in regards to dark matter candidates and baryogenesis generation\cite{cosmology, nucleosynthesis}. However, as with all fermionic dark matter candidates, sterile neutrino dark matter must satisfy the Tremaine-Gunn limit\cite{dm} which places absolute mass limits above \SI{0.5}{\kilo\electronvolt}\cite{whitepaper}. The stability of high mass sterile neutrinos must also be ensured by possessing weak Yukawa couplings in order to survive long enough to produce a meaningful impact on the universe's evolution. Thus, keV mass scales are highlighted as favourable regions for sterile neutrinos as both warm and cold dark matter candidates\cite{cdm}. It is worth noting also that low energy ($\textless$ \SI{100}{\mega\electronvolt}) RH neutrinos produce a vanishing rate of neutrinoless double beta decay even in a Majorana scenario\cite{rh}.

In particle physics, whilst sub-$\text{meV}$ mass scales are well excluded by standard oscillation experiments, the observed anomalies at SBLs are mostly consistent with a light sterile neutrino, which corresponds to the $\text{eV}$ mass scale.
\newpage
\subsection{Global Perspectives on Light Sterile Neutrinos}

As mentioned, the motivation surrounding light sterile neutrinos lies in anomalies of neutrino event rates at SBLs. Statistically significant anomalies have only ever been observed in \Pnue appearance channels at accelerators and \Pnue disappearance channels in reactor flux measurements and gallium source experiments.

\begin{figure}[htp]
\centering
\begin{subfigure}{.48\textwidth}
  \centering
  \includegraphics[width=\linewidth]{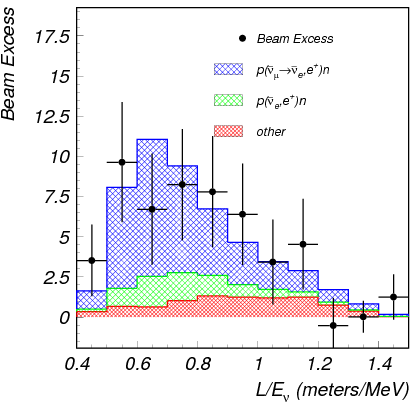}
  \caption{Final distribution of the CC \Pnue event excess (points) found in LSND\cite{lsnd}, compared with expectations (histograms) as a function of $L/E_{\nu}$ for energies in the range 20 \textless $E_e$\textless \SI{60}{\mega\electronvolt}.}
  \label{fig:LSND}
\end{subfigure}%
\hfill
\begin{subfigure}{.48\textwidth}
  \centering
  \includegraphics[width=\linewidth]{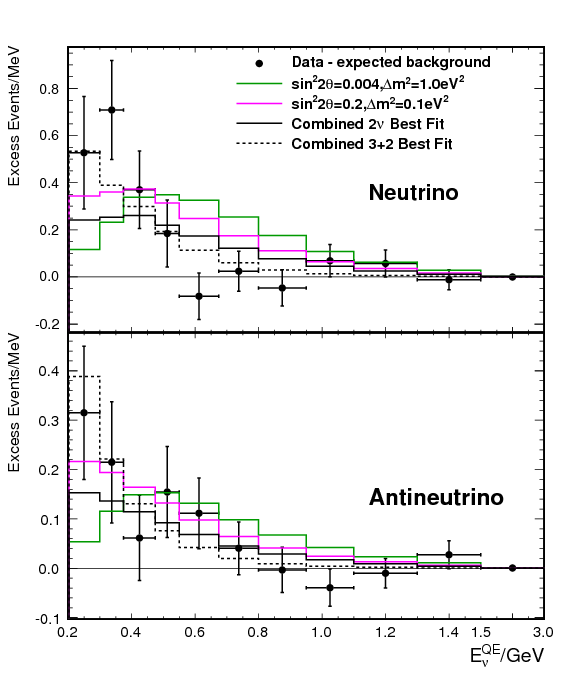}
  \caption{The CC quasi-elastic energy distributions of \Pnue event excesses found in MiniBooNE\cite{miniboone} in anti-neutrino mode (top) and neutrino mode (bottom), also showing example fits of oscillation parameters and the best $2\nu$ hypothesis fit for each mode.}
  \label{fig:miniboone}
\end{subfigure}
\caption{Observations of \Pnue CC event excesses found in fixed target accelerator experiments LSND and MiniBooNE.}
\label{fig:MiniBooNEExcess}
\end{figure}

The Liquid Scintillator Neutrino Detector (LSND) was the first accelerator based oscillation experiment to search for $\APnum \rightarrow \APnue$ electron appearance from a predominantly $\APnum$ beam. The neutrino beamline was established from decay-at-rest $\APmuon$ produced from interactions on a fixed target with a \SI{1}{\milli\ampere} \SI{798}{\mega\electronvolt} proton beam. Flux contributions from $\PKpm$ were negligible at this beam energy, with contributions from $\Pmuon$ and $\Pgpm$ mostly absorbed by iron shielding and the copper beam stop. The detector was a \SI{167}{\tonne} tank placed at \SI{30}{\meter} from the target, filled with a liquid scintillator mixture consisting of mineral oil doped with \SI[per-mode=symbol]{0.031}{\gram\per\litre} b-PBD. LSND employed \v{C}erenkov identification techniques, deploying 1220 8-inch photomultiplier tubes (PMTs) lined along the tank to identify light cones and neutron capture scintillation signals signifying Charged-Current (CC) inverse beta decay ($\APnue + \Pproton \rightarrow \Ppositron + \Pneutron$) interactions in the detector volume. After a 5 year run, LSND reported an excess of electron events at $L/E\sim$\SI{1}{\per\electronvolt\squared} corresponding to an oscillation probability of $(0.264 \pm 0.067 \pm 0.045)\%$ and an apparent $\Delta m^2$ signal in the range of \SIrange[range-phrase = --]{0.2}{10}{\electronvolt\squared}\cite{lsnd}, as shown in Figure \ref{fig:LSND}.

In 2002, the MiniBooNE experiment set out to investigate this excess further though with a longer baseline, a higher beam energy and with the ability to run in both neutrino and anti-neutrino modes. An \SI{8}{\giga\electronvolt} energy (\SI{8.89}{\giga\electronvolt\per\c} momentum) proton beam , established by Fermilab's Booster Accelerator, was aimed onto a beryllium target to produce mostly \PKpm, \Ppipm and \Pmupm. A magnetic focusing horn was deployed to filter out wrong sign particles, thus selecting through neutrino and anti-neutrino modes when pulsing with positive and negative polarities respectively. The filtered beam was then allowed to decay in flight through a \SI{50}{\meter} decay pipe, producing a predominantly \Pnum (\APnum) beam from decay chains of \Pgpp (\Pgpm) and \PKp (\PKm) with a small ($\sim1\%$) \Pnue (\APnue) contamination. The detector was placed at \SI{541}{\meter} from the target, consisting of a \SI{40}{\foot} sphere filled with \SI{806}{\tonne} of mineral oil and lined with 1520 8" PMTs, with an optically isolated outer veto region for cosmogenic event tagging. In the \SIrange[range-phrase = --]{0.2}{1.25}{\giga\electronvolt} region, excesses of $162.0 \pm 47.8$ ($3.4\sigma$) and $78.4 \pm 28.5$ ($2.8\sigma$) \Pe-like events were observed in neutrino and anti-neutrino modes respectively\cite{miniboone}, as shown in Figure \ref{fig:MiniBooNEExcess}b.

\begin{figure}[htp]
    \centering
    \includegraphics[width=0.9\linewidth]{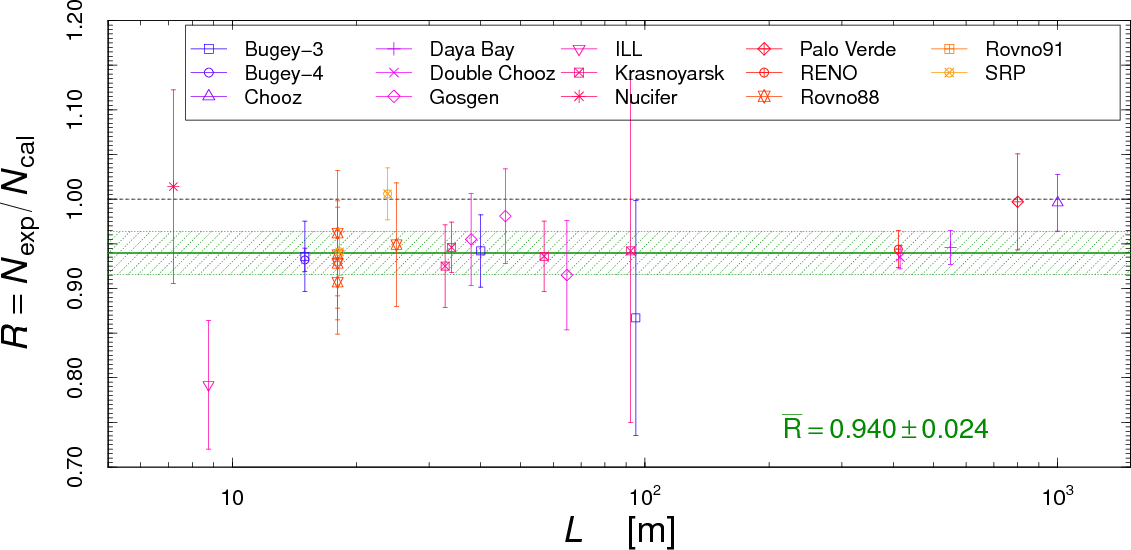}
    \caption{The observed to expected integrated \APnue event rate ratio as a function of detector distance as seen by various reactor experiments\cite{2017}.}
    \label{fig:reactor}
\end{figure}

In addition to these accelerator anomalies, a wide range of SBL $\APnue$ flux measurements made at nuclear reactor sites have observed a general deficit, dubbed the `reactor anomaly'. Until 2011, observed $\APnue$ fluxes resulting from $\beta$ decays of fission products in reactor cores were mostly consistent with a $3\Pneutrino$ mixing model. New flux predictions generated in preparation for Double Chooz\cite{reactorflux} pushed the averaged global observed to predicted event rate ratio of reactors from $0.976 \pm 0.024$ to $0.943 \pm 0.023$, which disfavours standard oscillations at $98.6\%$ CL\cite{reactoranomaly}. Figure \ref{fig:reactor} shows the observed to predicted event ratio for a collection of different reactor measurements.

As well, there exists the `gallium anomaly'. In solar neutrino observatories GALLEX and SAGE, the detectors were calibrated using neutrinos sourced from beta decay of \ce{^{51}Cr}. These detectors indirectly probed neutrino interactions from observation of the characteristic decay of \ce{^{71}Ge}, produced from inverse beta decay interactions upon the gallium target. The final average ratio of measured to predicted \Pnue flux in GALLEX and SAGE respectively is $0.93 \pm 0.08$\cite{gallex} and $0.87 \pm 0.11$\cite{sage}, to a combined statistical significance of 3.0$\sigma$\cite{gallium}.

A summary of recent 3+1 oscillation parameter fits to global data are available in Table \ref{tab:bestfits}. The oscillation region defined by anti-neutrino mode data from MiniBooNE possesses sufficient overlap with LSND's findings, though the MiniBooNE neutrino mode data is in significant tension to both of these\cite{Giunti2011a}. A 3+2 sterile model with large CP violation aids to resolve this, though this conflicts with limits set by big bang nucleosynthesis which allows for just one sterile neutrino at 95\% CL\cite{bbnn,bbn}. CP violating effects generated by non-standard interactions in a 3+1 scheme could also account for the disparity between MiniBooNE neutrino and anti-neutrino modes\cite{nsi}.


\begin{table}[htp]
	\centering
	\caption{Comparison of best fit points found by global data analyses for the set of 3+1 sterile oscillation parameters as made in 2013\cite{Kopp2013}, 2016\cite{Collin2016} and 2017\cite{2017}.}
	\label{tab:bestfits}
    \begin{tabular}{lcccc}\toprule
    ~& $\left|\Delta m^2_{41}\right|$ & $\left|U_{e4}\right|$ & $\left|U_{\mu 4}\right|$ & $\chi^2_{min}/dof$ ~           \\ \midrule
		2013 (Kopp et al.\cite{Kopp2013}) & 0.93 & 0.15 & 0.17 & 712/680 \\
		2016 (Collin et al.\cite{Collin2016})& 1.75 & 0.16 & 0.12 & 306.8/315\\
		2017 (S. Gariazzo et al.\cite{2017})& 1.70 & 0.16 & 0.12 & 594.8/579 \\ \bottomrule
    \end{tabular}
\end{table}

At this point in time, the supporting evidence for a light sterile neutrino is hazy and problematic in some regards. Particularly, tensions between MiniBooNE and LSND are high whilst the reactor anomaly could be as a result of inaccurate predictions\cite{reactorspectra}. The next generation of neutrino oscillation experiments should cross-check each of these anomalies with more powerful experimental techniques if more information on sterile neutrinos is to be provided. The SBN program's multi-detector setup will achieve just this, having been designed to provide a decisive refutation of the accelerator and reactor anomalies.
\newpage

\section{Overview of the SBN Program and Current Status}

Consisting of constituent experiments SBND\cite{lar1nd,lar1nd1} (previously known as LAr1-ND), MicroBooNE\cite{microboone} ($\mu$BooNE) and ICARUS\cite{icarus} (sometimes referred to as T600), Fermilab's Short-Baseline Neutrino program\cite{proposal} is a three detector program utilising state-of-the-art Liquid Argon Time Projection Chamber (LAr-TPC) technology in order to provide a comprehensive investigation into the possibility of light sterile neutrinos. The SBN program will utilise the onsite Booster Neutrino Beam (BNB) which has a well-understood flux, having already operated for over 10 years with the MiniBooNE and SciBooNE experiments.

\subsection{The Booster Neutrino Beam}

\begin{figure}[htp]
    \centering
    \includegraphics[width=\linewidth]{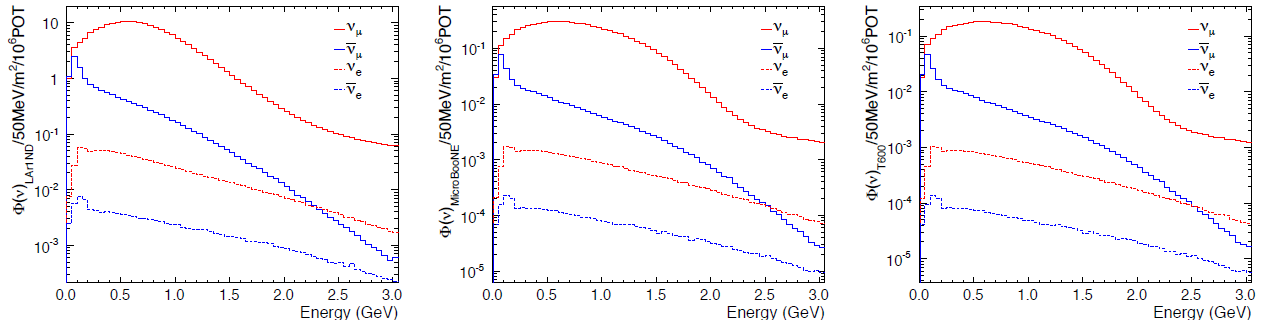}
    \caption{Expected neutrino flux distributions from the BNB as a function of energy at the three SBN detectors (Left) SBND (labelled as LAr1ND) (Middle) $\mu$BooNE (Right) ICARUS (labelled as T600)\cite{proposal}.}
    \label{fig:bnbflux}
\end{figure}

As mentioned, the BNB is well-characterised from accurate MiniBooNE simulations\cite{bnbflux} and so will operate with the same specifications as described for MiniBooNE in Section 2.2. The expected unoscillated fluxes at the three detectors are detailed in Figure \ref{fig:bnbflux}.

\subsection{LAr-TPC Detector Technology}

SBN is set to be the first formal instance of LAr-TPC deployment for a SBL oscillation search. The conceptual design of LAr-TPCs was first put forward in the late 1970's\cite{lartpc} but now seems to be the future of neutrino detection systems, provided it is practically feasible to deploy them at large scales. Since neutrino cross-sections are intrinsically so low, it is essential to be able to build detectors which operate stably with larger and larger target regions in order to overcome statistical restraints.

\begin{figure}[h!]
    \centering
    \includegraphics[width=0.7\linewidth]{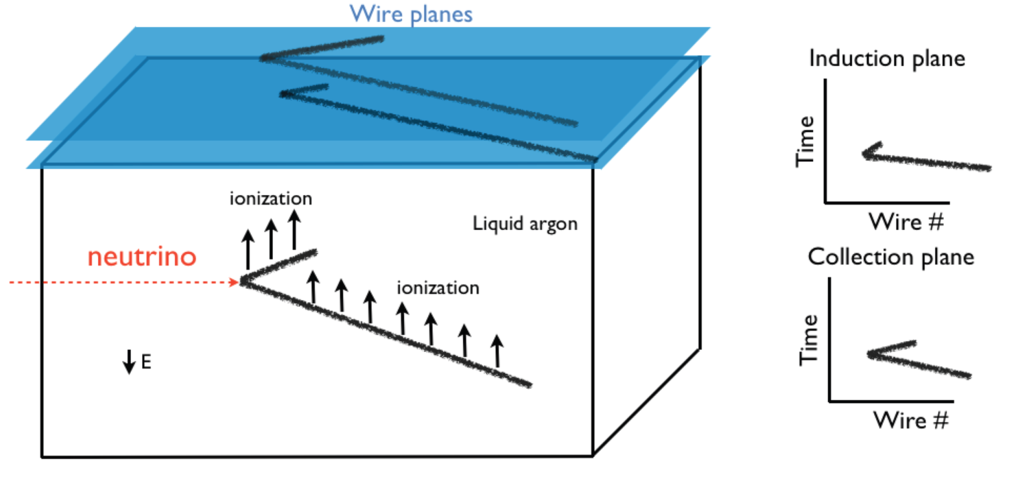}
    \caption{Principle of operation in LAr-TPCs. An applied transverse electric field drifts ionisation particles to wire planes. The deployment of multiple planes of wire arrays allows for multiple displays of the same event\cite{image}.}
    \label{fig:lartpc}
\end{figure}

Fundamentally, LAr-TPCs are tracking wire chambers coupled with a large volume of purified LAr which operates as both the nuclear target and ionising medium. Charged particles created from interactions on argon ionise the material as they traverse the detector, producing electrons and photons. Electromagnetic particles are subsequently drifted to planes of wire arrays by a strong uniform electric field where their energy manifests as a measurable electric potential on the wires. A 3D image of an event can then be constructed using wire number and drift time information, with calorimetric information deciphered from charge readout. PMTs are also deployed for the collection of scintillation light to be used as event triggers and for acquisition of temporal information. Figure \ref{fig:lartpc} offers a depiction of the operational principle of LAr-TPCs.

The intuitive event displays of LAr-TPC detectors seem to hark back to the days of bubble chambers albeit with much more powerful physics capabilities. The mm-level spatial resolution offered by LAr-TPCs provides fine-grained particle tracking and interaction reconstructions of high precision. Perhaps the most notable feature of LAr detectors is an adept separation of photon and electrons through a combination of topological and dE/dx calorimetric techniques. In \v{C}erenkov detectors, electromagnetic particles are indistinguishable, which stems from the fact that photons and electrons will shower at the same energy and produce identical rings. In LAr, electrons shower at the ionisation threshold energy whereas showers produced by high energy photons appear as a displaced vertex, as sufficient energy is required to be lost before pair production occurs. In the case where there is no topological gap observed, the dE/dx charge deposition at the onset of the shower can be used to distinguish particles\cite{mip}. An e-like shower is likely to lose energy at the rate of a Minimal Ionising Particle (MIP) (\SI{2.1}{\mega\electronvolt\per\centi\meter} in argon) whereas a $\gamma$-like shower is likely to lose energy at twice this rate, due to pair production. This feature aids to identify Neutral-Current (NC) \Ppizero production backgrounds which dominate \Pnue channels especially. Furthermore, the ability of resolving electromagnetic signals is integral to determining the event composition of the MiniBooNE excess to be electron-like or photon-like, as a result of appearance oscillations or consequential of some unknown background.

Significant efforts are focused on securing high LAr purity by limiting electronegative contaminants (mostly \ce{H2O}, \ce{O2}, \ce{CO2} etc.) which may attenuate ionisation tracks. It is possible to yield very long electron drift distances ($\sim$\SI{1}{\meter}) provided the LAr used is pure enough (contaminants kept at $<$ \SI{0.1}{\ppb}). As well, there exists a high ionisation yield ($\sim 6000$ $\Pelectron/\text{mm}$ for a MIP) and scintillation yield ($\sim5000$ $\gamma/\text{mm}$ for a MIP) giving greater data available for reconstruction\cite{mip2}. Argon, as a material, is also readily available and relatively low cost compared to the other noble gases, making it suitable for use in large quantities.

\begin{figure}[h!]
\begin{subfigure}{.5\linewidth}
\centering
\includegraphics[width=0.99\linewidth]{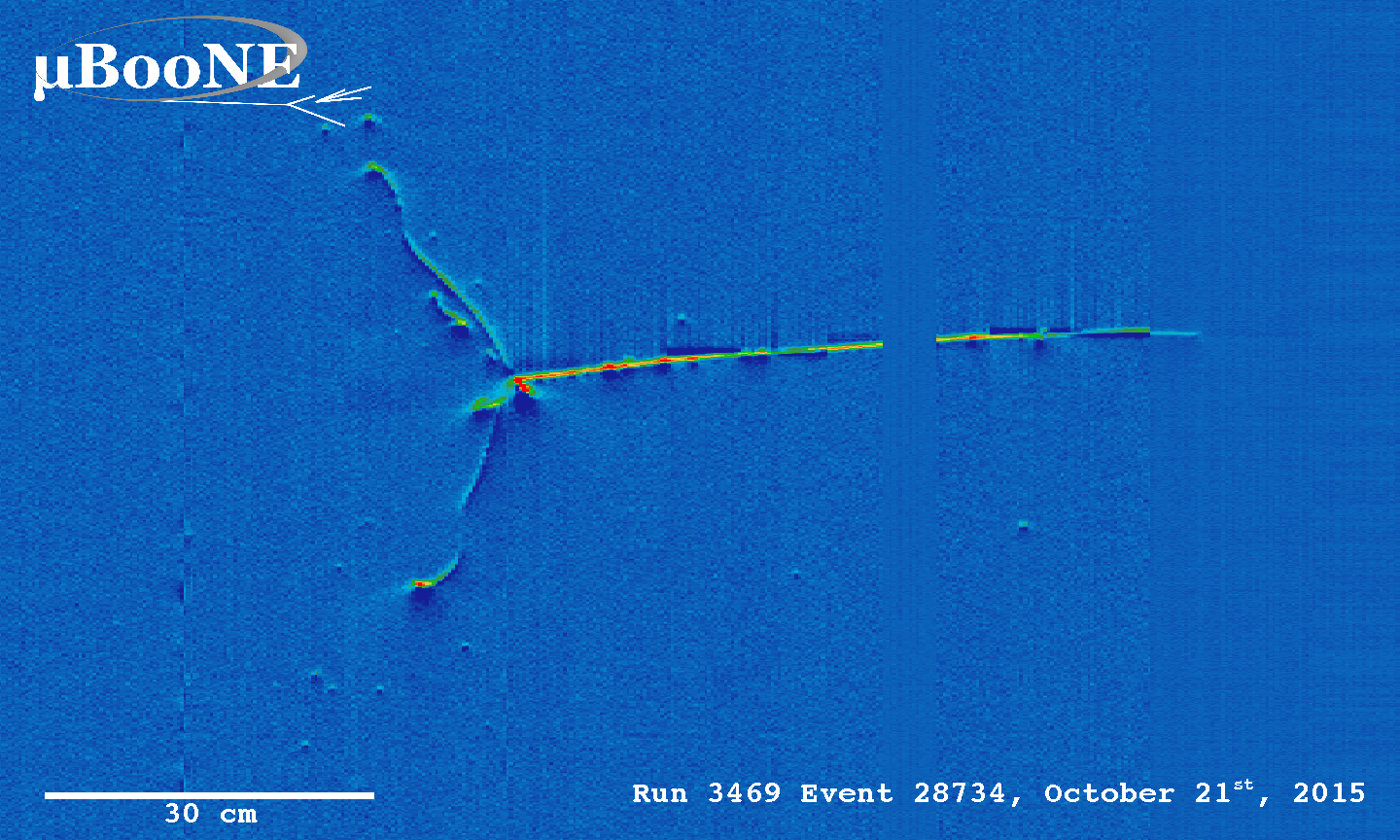}
\caption{U (induction) plane readout}
\label{fig:sub1}
\end{subfigure}%
\begin{subfigure}{.5\linewidth}
\centering
\includegraphics[width=0.99\linewidth]{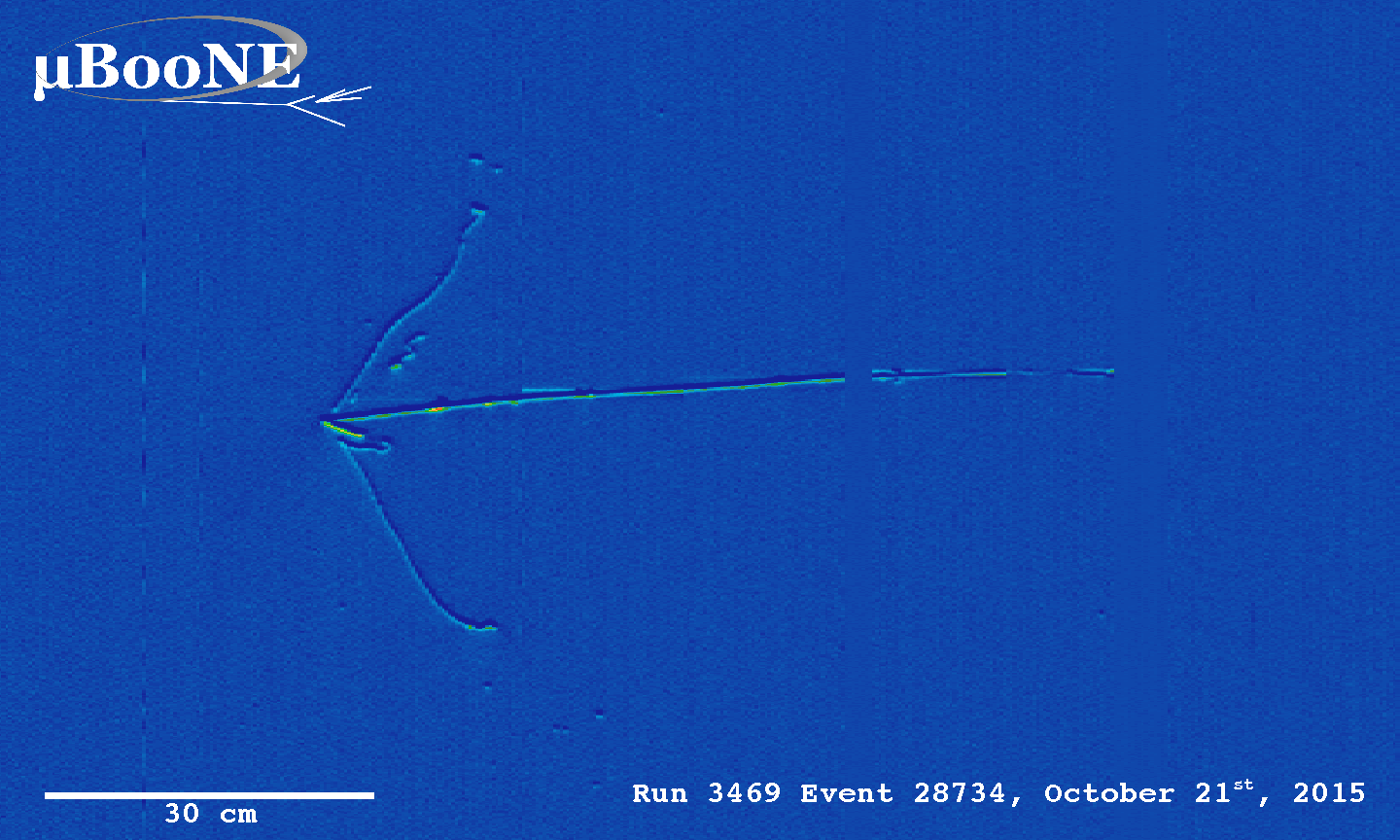}
\caption{V (induction) plane readout}
\label{fig:sub2}
\end{subfigure}\\[1ex]
\begin{subfigure}{\linewidth}
\centering
\includegraphics[width=.5\linewidth]{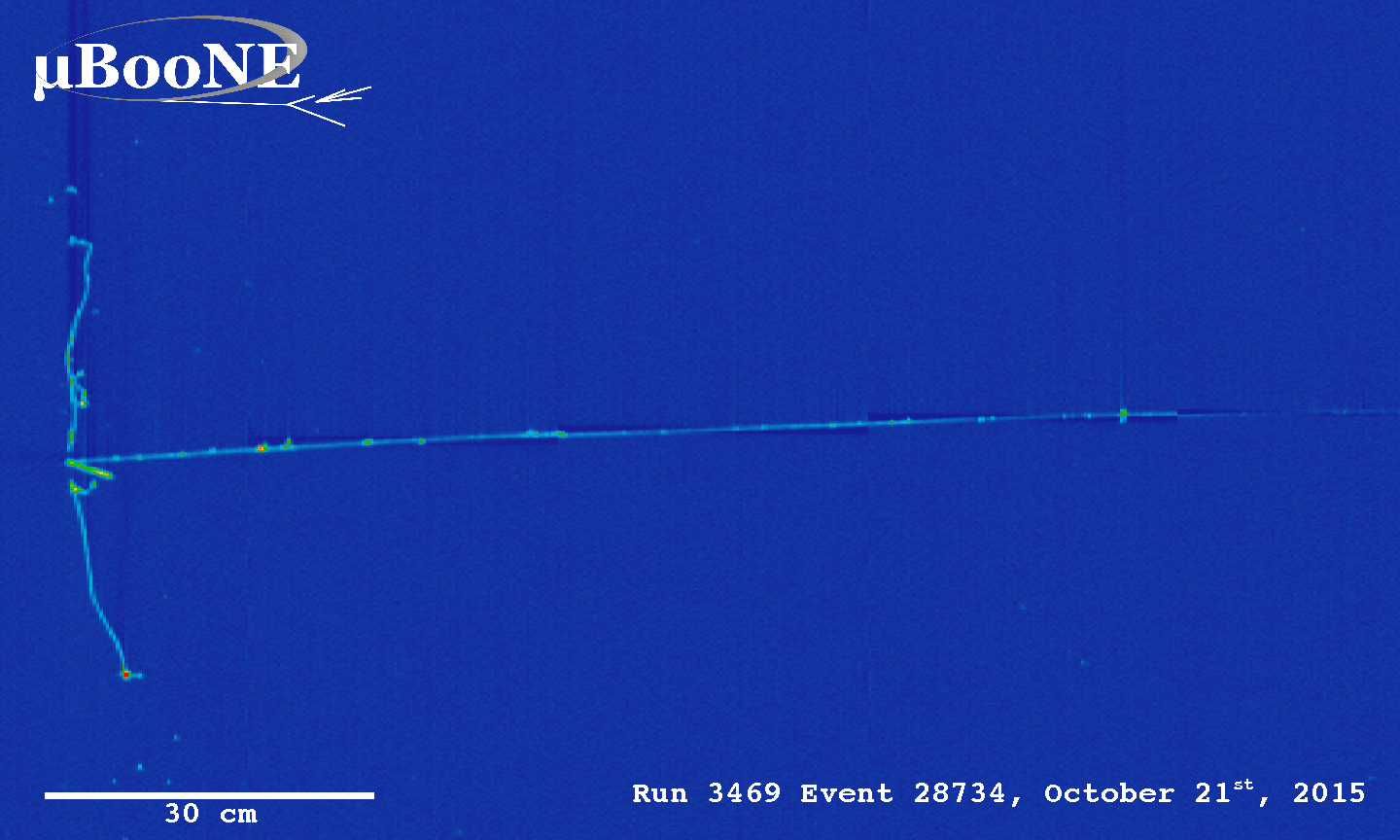}
\caption{Y (collection) plane readout}
\label{fig:sub3}
\end{subfigure}
\caption{Event displays of an example CC neutrino candidate event in $\mu$BooNE shown in each wire plane\cite{microbooneee}.}
\label{fig:test}
\end{figure}

A number of LAr tests have already been conducted. The ArgoNeuT project\cite{argoneut} deployed a \SI{750}{\litre} active volume module along the MINOS NuMI beamline and managed to demonstrate LAr's excellent electron-photon separation using calorimetry techniques\cite{argoneut1}. Serving as a preliminary test to DUNE, Fermilab's \SI{35}{\tonne} test was faced with complications\cite{protodune}. Still, the \SI{35}{\tonne} was still able to display a temporary stability of the full scale HV system at full voltage (\SI{120}{\kilo\volt}), establishing the expected DUNE drift field of \SI{500}{\volt\per\meter}. As well, the \SI{3}{\milli\second} electron lifetime necessary for useful tracking in DUNE was achieved, despite a contaminated LAr volume. The \SI{476}{\tonne} active mass ICARUS T-600 detector is still the largest LAr detector ever operated and performed quite stably from 2010-2013 with very high argon purity, observing atmospheric neutrinos and CERN's CNGS neutrino beam from a baseline of \SI{730}{\kilo\meter} at Grand Sasso National Laboratory\cite{icarus}. The \SI{87}{\tonne} $\mu$BooNE detector is the first formal deployment of an LAr-TPC for a SBL oscillation search and has been operating stably since October 2015. Figure \ref{fig:test} shows the event display of a candidate CC neutrino interaction in $\mu$BooNE. Ultimately, the goal in the next decade is to develop working LAr-TPCs with active LAr masses on scales of \SI{10}{\kilo\tonne} in preparation for the deployment of the DUNE far detector which is planned to comprise of four individual \SI{20}{\kilo\tonne} sub-detectors\cite{dune}. 

\subsection{The SBN Detection Systems}

The specifications of the SBN detectors are detailed in Table \ref{tab:specs}.

\begin{table}[h!]
\centering
\caption{Comparison of specifications of SBN's constituent experiments.}
\label{tab:specs}
    \begin{tabular}{cccc}\toprule
             & Baseline (m) & Total LAr Mass (tons) & Active LAr Mass (tons)  \\ \midrule
    SBND     & 110          &       210                & 112                  \\
    $\mu$BooNE & 470          & 170                   & 89                      \\
    ICARUS   & 600          &            760           & 476            \\\bottomrule
    \end{tabular}
\end{table}

SBND (Short-Baseline Near Detector) will operate close to the end of the decay pipe and will probe the mostly unoscillated spectrum. In addition to providing essential high-precision $\nu$-\ce{Ar} cross-section measurements, SBND will also search for the reactor anomaly by checking for disappearance of the intrinsic \Pnue content of the BNB. SBND is currently still in the design stage, with installation scheduled for late 2017. $\mu$BooNE commenced data acquisition in October 2015\cite{microbooneres} and is expected to accumulate \SI{6.6e20}{\POT} worth of data before SBND and ICARUS starts running. In addition to measuring $\nu$-\ce{Ar} cross-sections, the primary objective of $\mu$BooNE is to cross-check and resolve MiniBooNE's low energy electromagnetic excess by operating at a similar baseline and utilising the same beam. $\mu$BooNE also observes an off-axis flux of the Main Injector (NuMI) beam at an angle of \SI{110}{\milli\radian}. After decommissioning at Grand Sasso, ICARUS was transferred to CERN for refurbishment. ICARUS is still in the process of shipping to Fermilab, with installation also expected in late 2017. Overall, full operation of the SBN program is scheduled for mid 2018 with first run data available in spring 2019.

SBN is a multi-detector program, comprised of similar detectors, and will perform SBL oscillation searches at three different baselines by comparing the measured event spectra in each detector. In addition to a larger range of observable oscillation wavelengths, a further advantage of a multi-detector framework is the approximate cancelling of systematic uncertainties.
\subsection{Current Work and Motivation of Study}

\begin{figure}[htp]
\centering
\begin{subfigure}{.48\textwidth}
  \centering
      \includegraphics[width=1.0\linewidth]{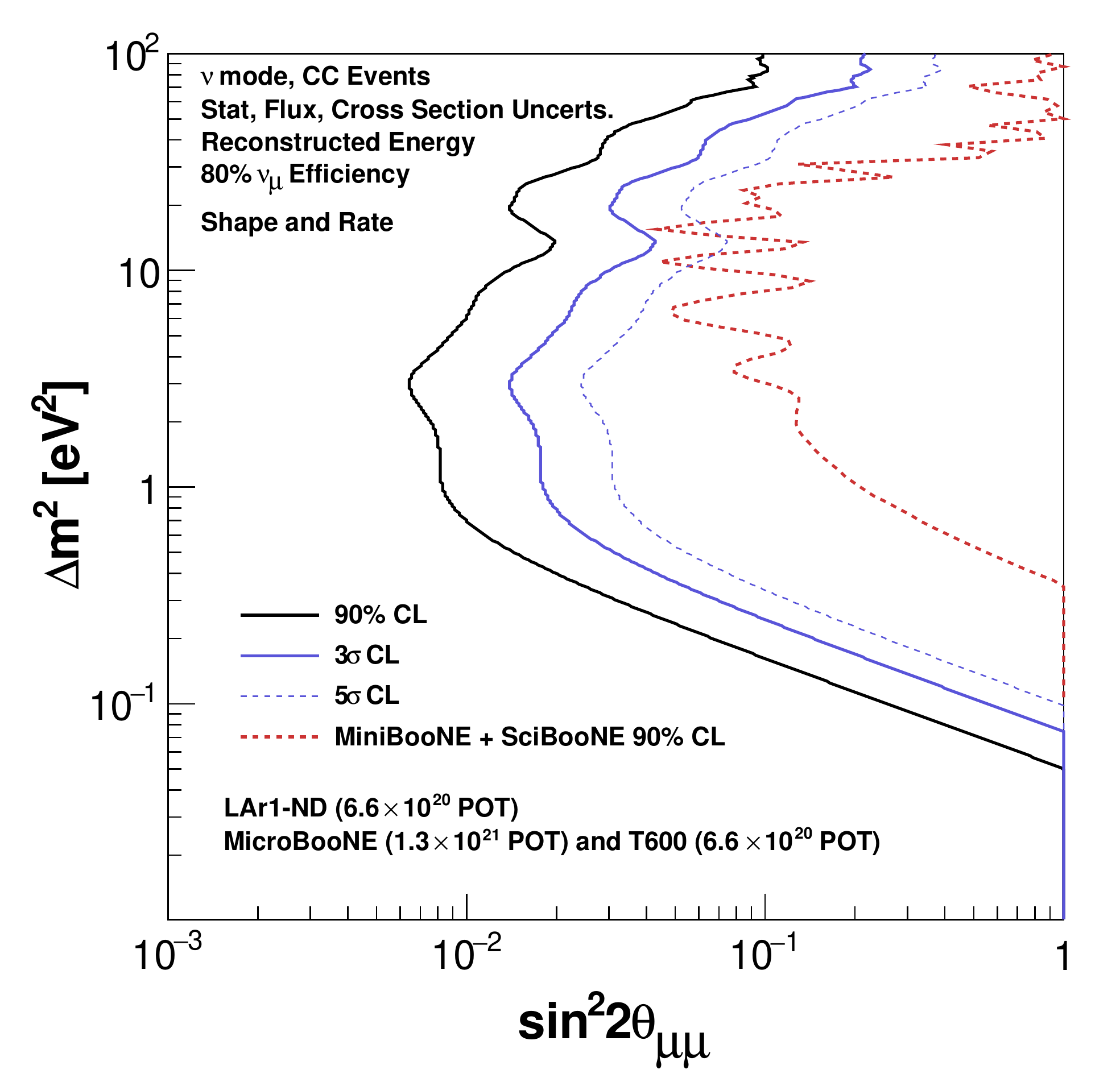}
    \caption{Predicted sensitivity of the SBN program to oscillation parameters $\Delta m^2 $ and $\sin^2(2\theta_{\mu\mu})$ from a CC \Pnum analysis, showing relation to parameter space covered by combining MiniBooNE and SciBooNE data\cite{scibooneminiboone}.}
    \label{fig:sbnmumode}
\end{subfigure}%
\hfill
\begin{subfigure}{.48\textwidth}
  \centering
   \includegraphics[width=1.\linewidth]{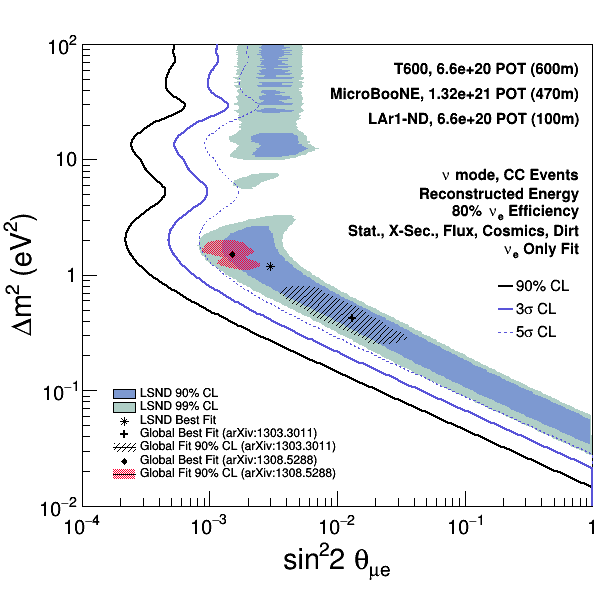}
    \caption{Predicted sensitivity of the SBN program to oscillation parameters $\Delta m^2 $and$ \sin^2(2\theta_{\mu e})$ from a CC \Pnue appearance analysis, showing relation to space already covered by LSND and favoured by two global fits\cite{Kopp2013, giunti2013}.}
    \label{fig:sbnemode}
\end{subfigure}
\caption{SBN sensitivity plots as produced by original proposal paper\cite{proposal}}
\label{fig:sbnsens}
\end{figure}

As shown in Figure \ref{fig:sbnsens}, experimental sensitivities of the SBN program for separate channels have already been produced using a flux$\times$cross-section integration method with oscillations implicitly encoded\cite{proposal}. In addition, Ref. \cite{sbn2} presents SBN sensitivity plots on an event-by-event basis with an explicit focus on 3+2 parameters.

As of yet, there have been no efforts to combine \Pnue appearance and \Pnum disappearance channels nor to find their sensitivity representation in terms of 3+1 mixing angles. This report sets out to do so, using an event-by-event strategy to enable a more realistic reconstruction emulation.
\newpage
\section{Analysis Procedure}

\subsection{Event Generation and Detector Emulation}

This analysis study was based around manipulation of input data files produced from a fast Monte Carlo simulation. The interactions of neutrinos on argon were facilitated by the GENIE event generator, then a parametrised reconstruction was applied using geometries of a `toy' detector.

Events in the detector, after applying pseudo-reconstruction smearing, were classified by the produced track's physical containment within the detector, their reconstructed energy and flavour of the reconstructed lepton. The efficiencies and resolutions applied are detailed in Table \ref{tab:efficiency}. Useful events, which provide enough data for a meaningful reconstruction, were given a `good' event tag, warranted by one or more final state particles in the event with kinetic energy exceeding \SI{50}{\mega\electronvolt}. The energies of the hadronic and leptonic components of the final state interaction products were smeared individually around their true value by the appropriate energy resolution with a Gaussian function, and then combined to produce a reconstructed neutrino energy. The resulting CC \Pnue efficiency corresponds to $\sim$80\% with a NC rejection factor of 99\%.

\begin{table}[h!]
\centering
\caption{Applied efficiencies and resolutions in the pseudo-reconstruction stage.}
\label{tab:efficiency}
\begin{tabular}{lll}
\toprule
& Event Type & Applied Value \\ \midrule
Angular Resolution & Electron & \ang{1} \\
 & Muon & \ang{1} \\
 & Hadronic System & \ang{10} \\ \midrule
Energy Resolution & Stopping Muon & 3\% \\
 & Exiting Muon & 15\%\cite{lbne} \\
 & Electron & 1\%/$\sqrt{E}$ + 1\%\cite{icarus} \\
 & Hadronic System & 30\%/$\sqrt{E}$\cite{lbne} \\ \midrule
Signal Acceptance & Electron & 90\% \\
 & Muon & 100\% \\ \midrule
Background Rejection & e-like (\Ppizero, \Pgamma) & 95\% \\
 & $\mu$-like (\Ppiplus, \Ppiminus) & 99\% \\\bottomrule
\end{tabular}
\end{table}

A tabulated data sample consisting of 1 million neutrino and anti-neutrino events for a `toy' detector modelled on the dimensions and specifications of $\mu$BooNE was created for analysis. Another data sample was created, swapping the \Pnum content of the BNB for \Pnue and vice versa, as a tool for adding on appearing \Pnue events.

\subsection{Simulation of SBL Oscillation Signatures}

SBN's' $L/E$ scale of interest was considered different enough to oscillations facilitated by the standard oscillation parameters such that a two flavour approximation oscillation model was adopted, considering only transitions of \Pnue and \Pnum generated by the $\Delta m^2_{41}$ mass splitting. This formalism is justified by LAr's poor sensitivity to \Pnut, such that $\left|U_{\tau i}\right|^2 \approx 0$ is assumed for all $i$, and the inherent undetectability of $\Pnu_s$. It is also presumed that there is negligible presence of \Pnut and $\Pnu_s$ intrinsic of the BNB which may undergo flavour transitions to appear as \Pnue or \Pnum. Equation \ref{eq:oscc} gives the full general transition probability function in natural units.

    \begin{multline} \label{eq:oscc}
	P_{\nu_{\alpha} \rightarrow \nu_{\beta}} = \delta_{\alpha\beta} - 4 \sum_{i > j}\text{Re}\left(U^*_{\alpha i}U_{\beta i}U_{\alpha j}U^*_{\beta j}\right)\sin^2\left(\frac{\Delta m^2_{ij}L}{4E}\right) \\+ 2\sum_{i>j}\text{Im}\left(U^*_{\alpha i}U_{\beta i}U_{\alpha j}U^*_{\beta j}\right)\sin\left(\frac{\Delta m^2_{ij}L}{2E}\right)
	\end{multline}
	
	where parameters $U_{ij}$ correspond to elements of the PMNS matrix, $\Delta m^2_{ij}$ is the mass squared splitting between states $i$ and $j$, $L$ is the distance along the neutrino beam line and $E$ is the neutrino energy.
	
    The simplified SBL oscillation formula applied throughout this investigation is yielded by selecting the oscillation mode where $i=4$ and $j=1$ and by assuming a two flavour approximation model is sufficient to describe SBL oscillations. Through this, the PMNS matrix reduces to a \num{2 x 2} matrix describing the mixing of \Pnum and \Pnue flavour eigenstates in the basis of the 1st and 4th mass eigenstates. Furthermore, through enforcing unitary of the reduced mixing matrix, SBL oscillation amplitudes can be described solely by the mixing elements associated with the 4th mass eigenstate, such that $\left|U_{\alpha1}\right|^2 \simeq \left(\delta_{\alpha\beta} - \left|U_{\beta4}\right|^2\right)$. The resulting 3+1 SBL oscillation formula is given in Equation \ref{eq:2flav}.

\begin{equation} \label{eq:2flav}
\Rightarrow P^{3+1}_{\nu_\alpha \rightarrow \nu_\beta} \simeq \delta_{\alpha\beta} - 4\left|U_{\alpha4}\right|^2\left(\delta_{\alpha\beta} - \left|U_{\beta4}\right|^2\right)\sin^2\left(\frac{1.27\Delta m^2_{41}L}{E}\right)
\end{equation}
where the argument of the oscillating sinusoid is rewritten in more convenient units, so that $\Delta m^2_{41}$ is in \si{\electronvolt\squared}, $L$ is in \si{\kilo\meter} and $E$ is in \si{\giga\electronvolt}.

Since the ordering of the mass eigenstate rotations given by the PMNS matrix is arbitrary, it's parametrisation is a matter of taste. This study applied a parametrisation of $U$ given by Equation \ref{eq:param}\cite{rebel}.

\begin{equation}\label{eq:param}
U = R_{34}(\theta_{34})R_{24}(\theta_{24},\delta_2)R_{14}(\theta_{14})R_{23}(\theta_{23})R_{13}(\theta_{13},\delta_1)
\end{equation}
where $R_{ij}$ are rotation matrices between states $i$ and $j$ parametrised by mixing angles $\theta_{ij}$ and phases $\delta_k$.
The rotation matrices of $U$ are of a form given by Equation \ref{eq:form}. 

\begin{equation} \label{eq:form}
    R^{pq}_{ij}(\theta_{ij},\delta_k) =
\begin{cases}
    \cos\theta_{ij},& p = q = i \quad \text{or}\quad p = q = j, \\
    1,& p = q \neq i \quad \text{and}\quad p = q \neq j, \\
    \sin\theta_{ij}e^{-i\delta_{k}}, & p = i \quad\text{and}\quad q = j, \\
    -\sin\theta_{ij}e^{i\delta_k}, & p = j \quad\text{and}\quad q = i, \\
    0, & \text{otherwise.}
\end{cases}
\end{equation}
where $p$ and $q$ are the rows and columns of $R_{ij}$ respectively.

Therefore, the elements of the PMNS matrix relevant to SBL oscillations of \Pnum and \Pnue can be found, as given in Equations \ref{eq:e4} and \ref{eq:mu4}.

\begin{equation}\label{eq:e4}
\left|U_{e4}\right|^2 = \sin^2\theta_{14}
\end{equation}

\begin{equation}\label{eq:mu4}
\left|U_{\mu4}\right|^2 = \cos^2\theta_{14}\sin^2\theta_{24}
\end{equation}

The oscillation channel amplitudes may be more conveniently represented by single parameter functions, as given by Equations \ref{eq:alphabeta} and \ref{eq:alphaalpha}.

\begin{equation}\label{eq:alphabeta}
\sin^2\left(2\theta_{\alpha\beta}\right) = 4\left|U_{\alpha4}\right|^2\left|U_{\beta4}\right|^2, \qquad \alpha \neq \beta
\end{equation}

\begin{equation}\label{eq:alphaalpha}
\sin^2\left(2\theta_{\alpha\alpha}\right) = 4\left|U_{\alpha4}\right|^2\left(1- \left|U_{\alpha4}\right|^2\right)
\end{equation}

Since a sterile neutrino is invisible to both the \PWpm and \PZ bosons, oscillations are expected in both the CC and NC channels. This analysis is a study of sensitivities to SBL oscillations from studying CC events only, therefore NC oscillation effects are neglected.

\subsection{Analysis Strategy}

The basic analysis procedure was to select \Pnum or \Pnue events by their reconstructed flavour and interaction type and bin them by their reconstructed energy. The bulk of the data stemmed from the \Pnum content of the BNB whilst the \Pnue spectra arose in majority from the 1\% intrinsic \Pnue content. With oscillations applied, visible disappearance was expected in the \Pnum channel whilst both appearance and disappearance was expected in the \Pnue channel, though with appearance dominating due to the high \Pnum content of the BNB. Any \Pnum appearance contributions were regarded as negligible considering the already small \Pnue intrinsic content of the BNB and suppressed oscillation probability. The signal and backgrounds were postulated to be well separated, given the excellent particle identification capabilities of LAr which renders both NC and misidentification backgrounds to be small.

Encoding of oscillations and analysis were both performed exclusively through the scientific software framework ROOT in C++. In creating energy spectra, neutrino events were selected from the data chain according to their reconstructed flavour and added to a 1D histogram by their reconstructed energy. Tabulated data available in Ref. \cite{proposal} for the total expected event rates in SBND and $\mu$BooNE were used to normalise the statistics appropriately. The expected event rates for ICARUS are not yet available and so were approximated by scaling $\mu$BooNE event rates up by the ratio of their active LAr masses and down by the ratio of their detector baselines squared. The inclusion of the SBL disappearance oscillation channel was performed simply by weighting the individual event binning by the disappearance oscillation probabilities given by Equation \ref{eq:2flav}. Accounting for appearance channel effects is a more complicated process and is described in Section 6.1.

\begin{figure}[h!]
    \centering
    \includegraphics[width=0.7\linewidth]{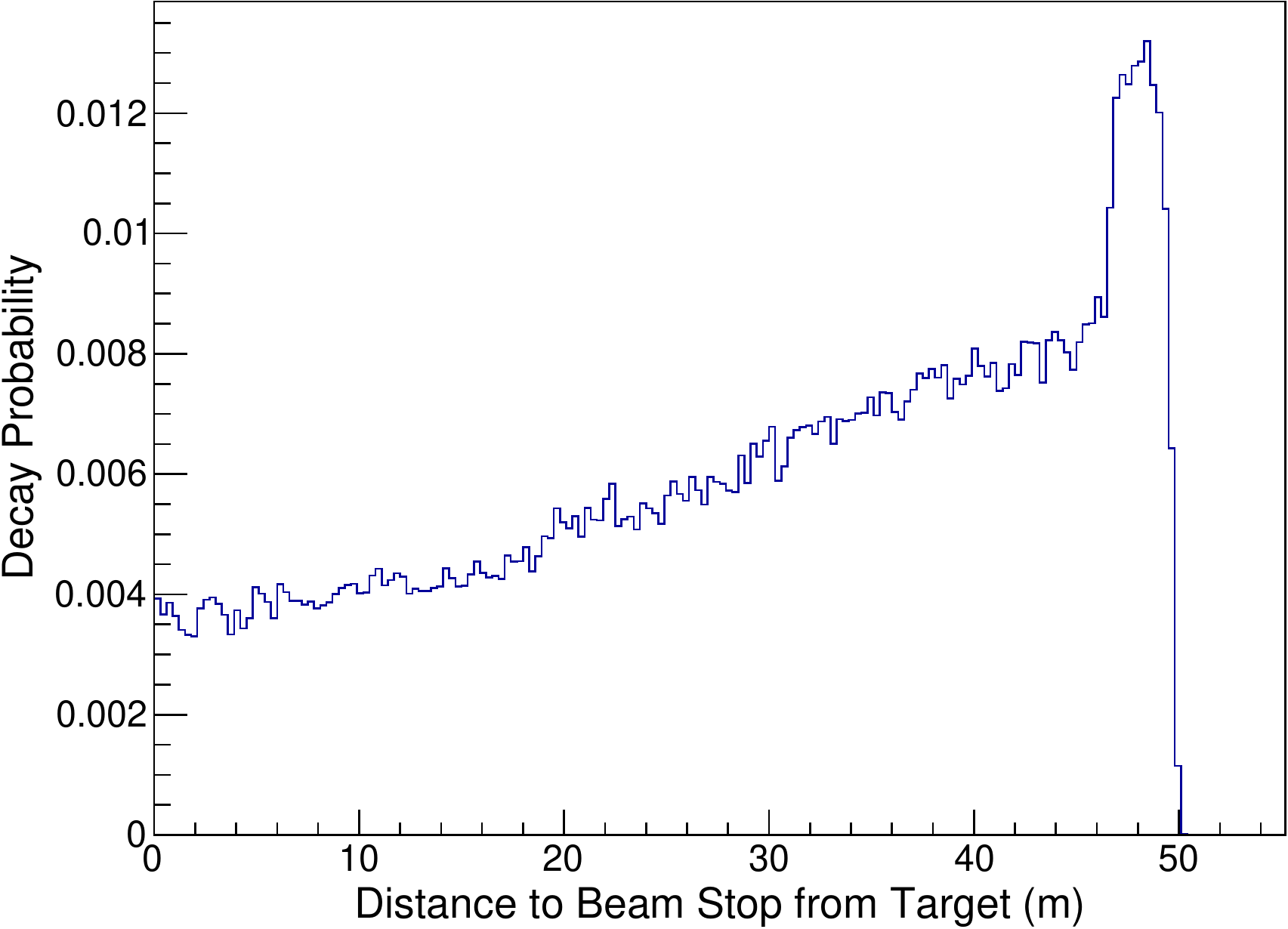}
    \caption{Probability distribution of decay points of primary mesons along the \SI{50}{\meter} BNB decay pipe. (Study credited to Dr. Dom Brailsford of Lancaster University).}
    \label{fig:decaydist}
\end{figure}

The distance that a neutrino propagates before interaction is dependent on the lifetime and momentum of it's parent particles traversing through the BNB's \SI{50}{\meter} decay pipe. This effect creates a distribution of effective baselines and hence a distribution of effective oscillation probabilities visible at a particular detector. Though expected to be of minor significance, this occurrence was speculated to smear out oscillation signals and therefore reduce overall SBN sensitivities. Figure \ref{fig:decaydist} shows the probability distribution of decay points in the decay pipe produced from a Monte Carlo simulation of Booster target interactions. 

Through the Geant4 toolkit, the simulation incorporates the beam components, electromagnetic fields and the physics of propagation and interaction in recreating hadroproduction processes from the Booster Accelerator's primary protons acting on the beryllium target. The focusing and then subsequent decay of secondary hadrons into the decay pipe is then considered, where the production point of the newly created neutrinos is plotted. The distribution of decay points peaks within the first \SI{4}{\meter} of target interaction and reduces steadily after. The beam stop is then assumed to be completely efficient in absorbing any remaining charged particles in the decay pipe. For each event in each detector, the effective oscillation baseline used is equal to the detector distance subtracted by the distance of a random decay point selected in the above probability distribution. The significance of this effect on sensitivities is discussed in Appendix B. Furthermore, the energy applied in the oscillation weightings is the true neutrino energy as opposed to the reconstructed energy.

The two event ratios each of the far detectors to the near detector are illustrated throughout this study, a technique inherited from sterile searches in MINOS\cite{minos}. That is, the energy spectra of $\mu$BooNE divided by the spectra of SBND and the energy spectra of ICARUS divided by the spectra of SBND. This technique is considered to be not only useful for characterising the degree of oscillation across the detectors, but also to cancel out uncertainties associated with cross-section physics.

The experimental sensitivities, or the degree of the statistical significance of SBL anomalies generated by a particular set of oscillation parameters, were characterised by a $\chi^2$ value comparing a sterile outlook to a null unoscillated hypothesis for each of the event ratios. The general $\chi^2$ formula for a particular event ratio to the near detector is defined by Equation \ref{eq:chisq}.

\begin{equation}\label{eq:chisq}
\chi^2\left(\Delta m^2_{41}, \theta_{14}, \theta_{24}\right) = \sum^N_{i}\left(\frac{R^{\text{Osc.}}_i\left(\Delta m^2_{41}, \theta_{14}, \theta_{24}\right) - R^{\text{Unosc.}}_i}{\sigma_i \left(\Delta m^2_{41}, \theta_{14},\theta_{24}\right)}\right)^2
\end{equation}
where $R^{\text{Osc.}}_i$ and $R^{\text{Unosc.}}_i$ are the contents of bin $i$ of the event ratio spectra with oscillations and no oscillations respectively. $\sigma_i$ is the uncorrelated error given by the oscillated ratio of a particular bin $i$.

The error $\sigma$ for a energy bin $i$ was calculated by propagating Poisson errors, as given in Equation \ref{eq:error}.

\begin{equation}\label{eq:error}
\sigma_i\left(\Delta m^2_{41}, ...\right) = R_i\left(\Delta m^2_{41}, ...\right)\sqrt{\frac{1}{N^{\text{FAR}}_i\left(\Delta m^2_{41}, ...\right)} + \frac{1}{N^{\text{SBND}}_i\left(\Delta m^2_{41}, ...\right)}}
\end{equation}
where $N_i$ is the content of energy bin $i$ for a particular detector.

The two $\chi^2$ values obtained from the two event ratios were simply added to produce a total $\chi^2$, as shown in Equation \ref{eq:add}, to characterise the full program's multi-detector statistical significance of applied SBL oscillations.

\begin{equation} \label{eq:add}
\sum_{i}\chi^2_i\left(\Delta m^2_{41}, ...\right) = \chi^2_{\mu\text{BooNE/SBND}}\left(\Delta m^2_{41}, ...\right) + \chi^2_{\text{ICARUS/SBND}}\left(\Delta m^2_{41}, ...\right)
\end{equation}

These $\chi^2$ formulae were later applied to construct 2-dimensional sensitivities in parameter space, as demonstrated in Sections 5.2 and 6.2.

\section{Muon Disappearance Analysis}
The first stage of analysis is to demonstrate the multi-detector sensitivity to SBL \Pnum disappearance oscillations.
\subsection{Energy Spectra Predictions}
    
The \Pnum energy spectra was drawn by combining all `good' events which reconstruct a muon, whilst highlighting the background contribution from true NC events which are reconstructed, and thus misidentified, as CC events. Oscillations are applied to each CC event by weighting the binning with the disappearance probability as given by Equation \ref{eq:mudisprob}.

\begin{equation}\label{eq:mudisprob}
P_{\nu_\mu \rightarrow \nu_\mu} = 1 - \sin^2\left(2\theta_{\mu\mu}\right)\sin^2\left(\frac{1.27\Delta m^2_{41}L}{E}\right)
\end{equation}

\begin{figure}[ht!]
    \centering
    \includegraphics[width=0.70\linewidth]{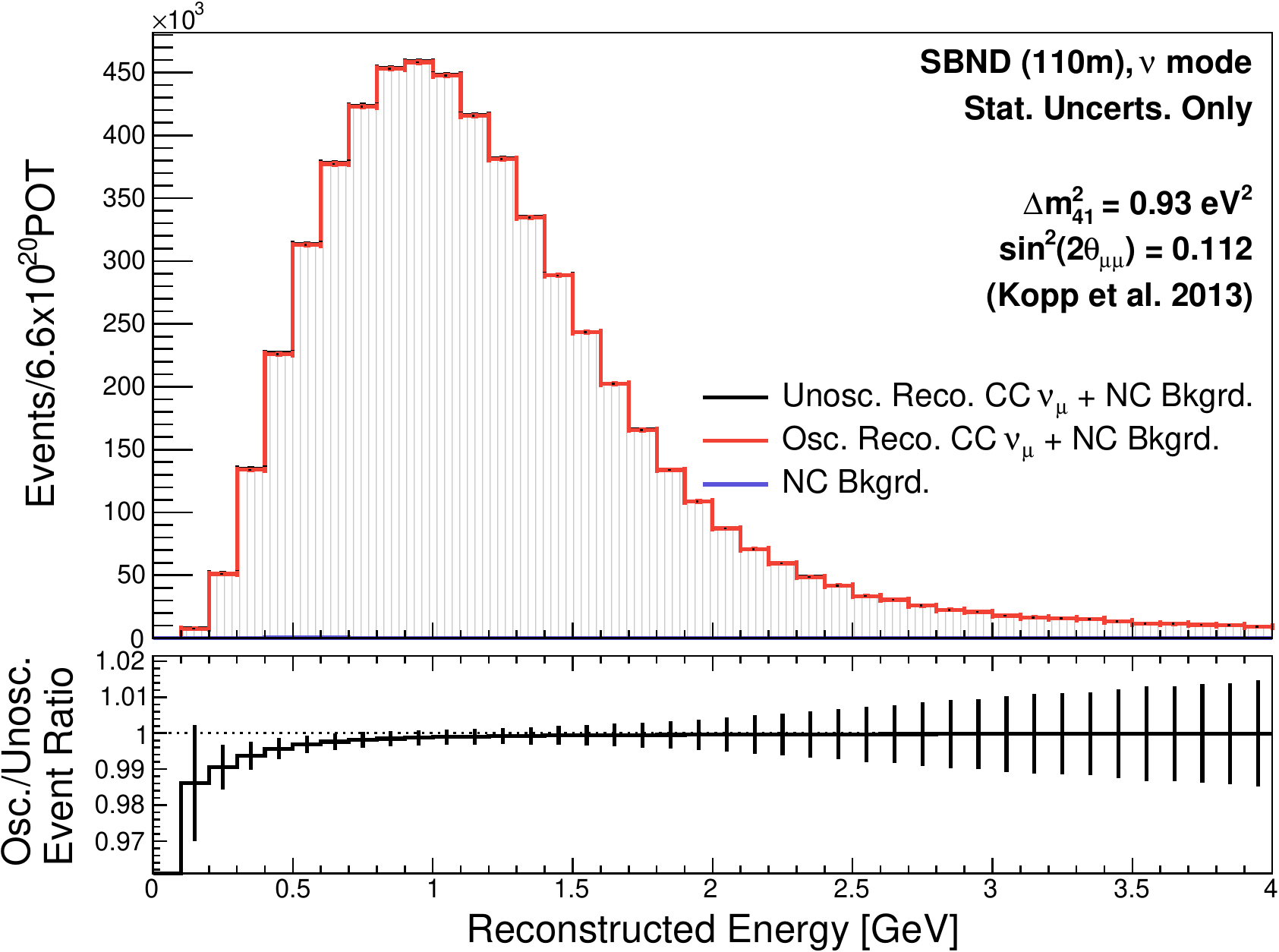}
    \caption{(Top) CC \Pnum event distribution in SBND as a function of reconstructed neutrino energy, applying oscillations generated by Kopp et al. 3+1 best fit parameters. The resulting disappearance spectra (red), unoscillated spectra (black) and NC background (blue) are presented with statistical uncertainties only. (Bottom) Ratio of oscillated to unoscillated spectra, parallelling the above distribution.}
    \label{fig:mudissbnd}
\end{figure}

\begin{figure}[hb!]
    \centering
    \includegraphics[width=0.70\linewidth]{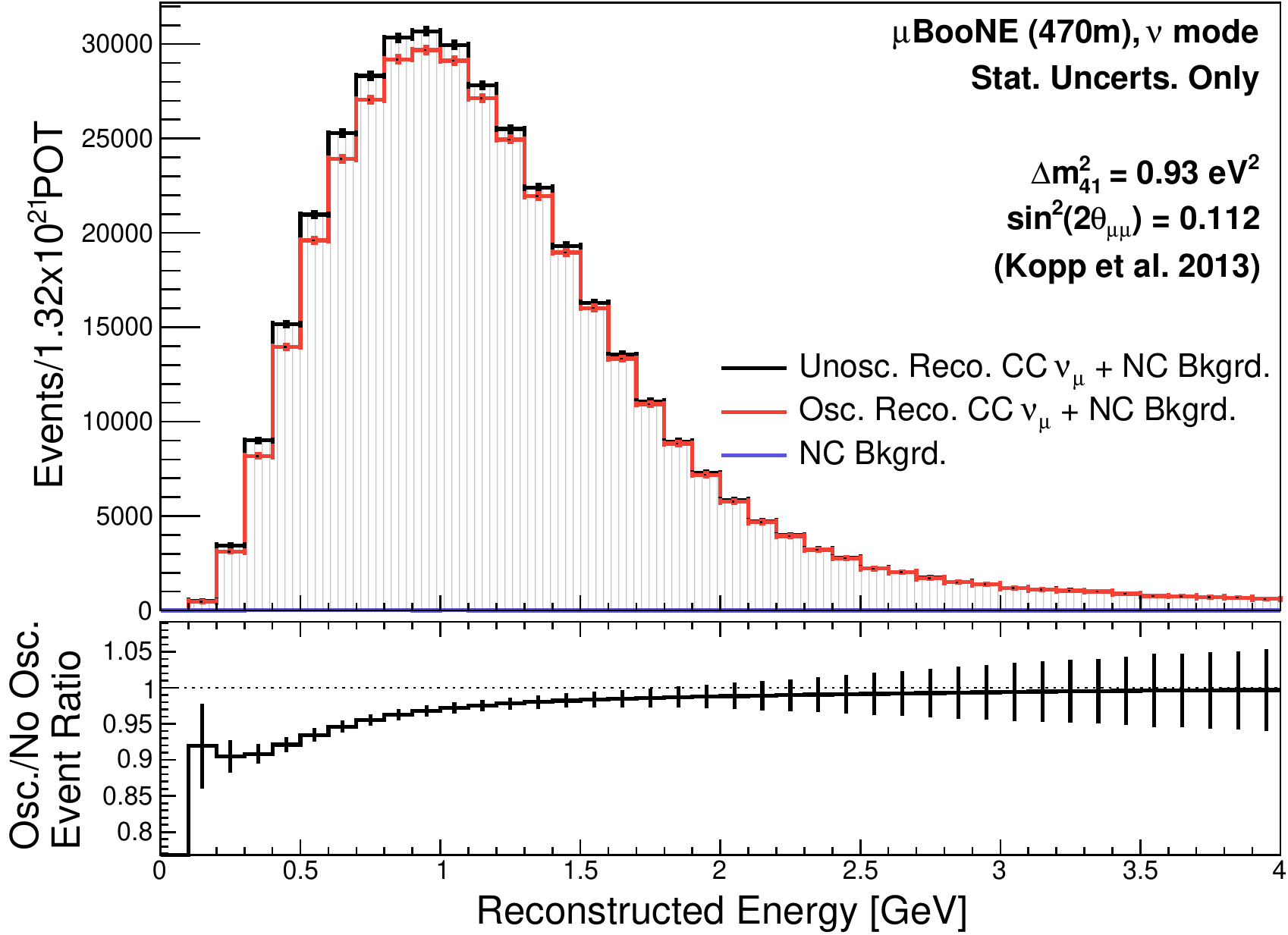}
    \caption{Same as Figure \ref{fig:mudissbnd} except with event rates observed at $\mu$BooNE.}
    \label{fig:mudismicroboone}
\end{figure}

\begin{figure}[h!]
    \centering
    \includegraphics[width=0.716\linewidth]{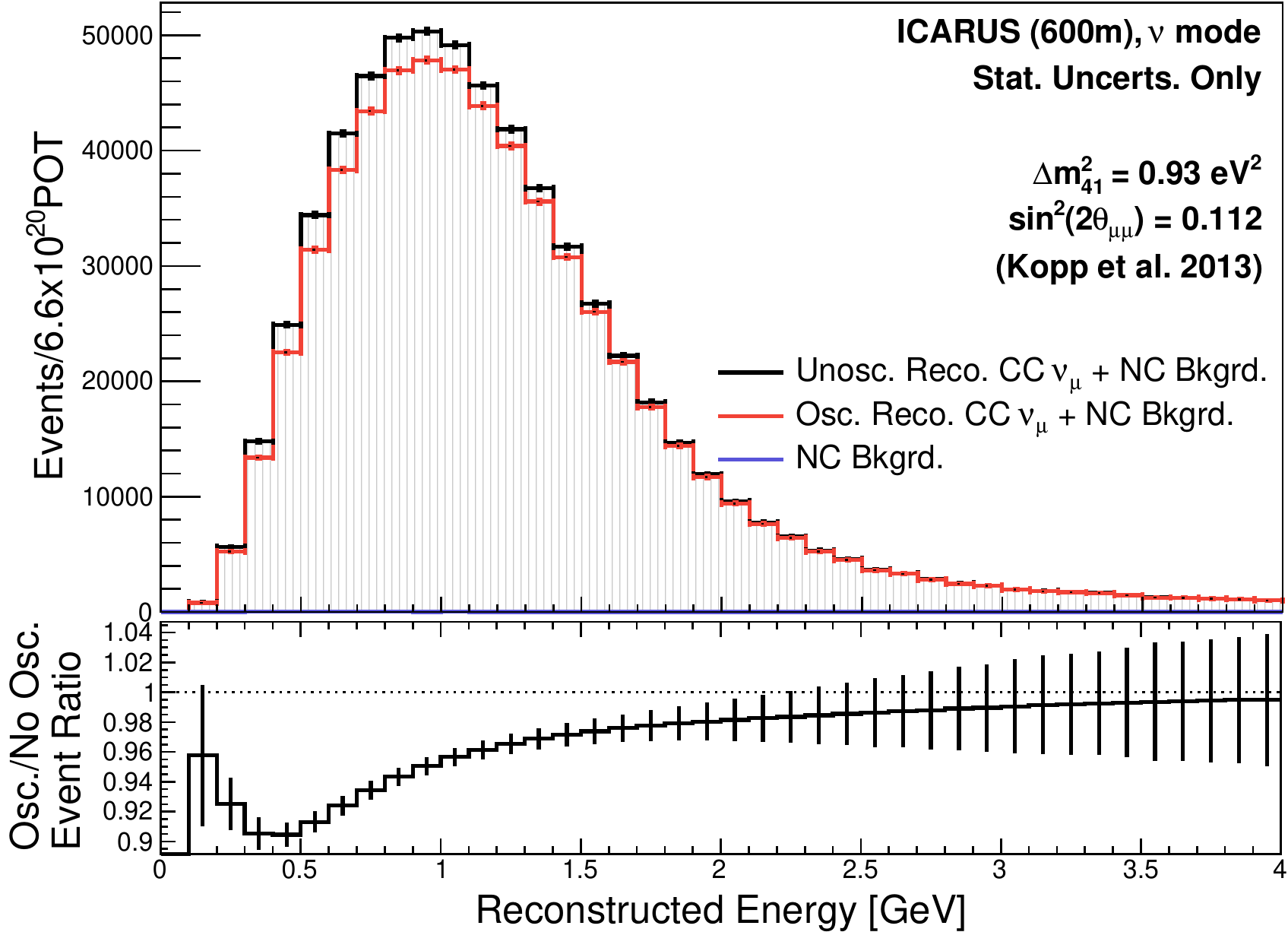}
    \caption{Same as Figure \ref{fig:mudissbnd} except with event rates observed at ICARUS.}
    \label{fig:mudisicarus}
\end{figure}

\begin{figure}[h!]
    \centering
    \includegraphics[width=0.76\linewidth]{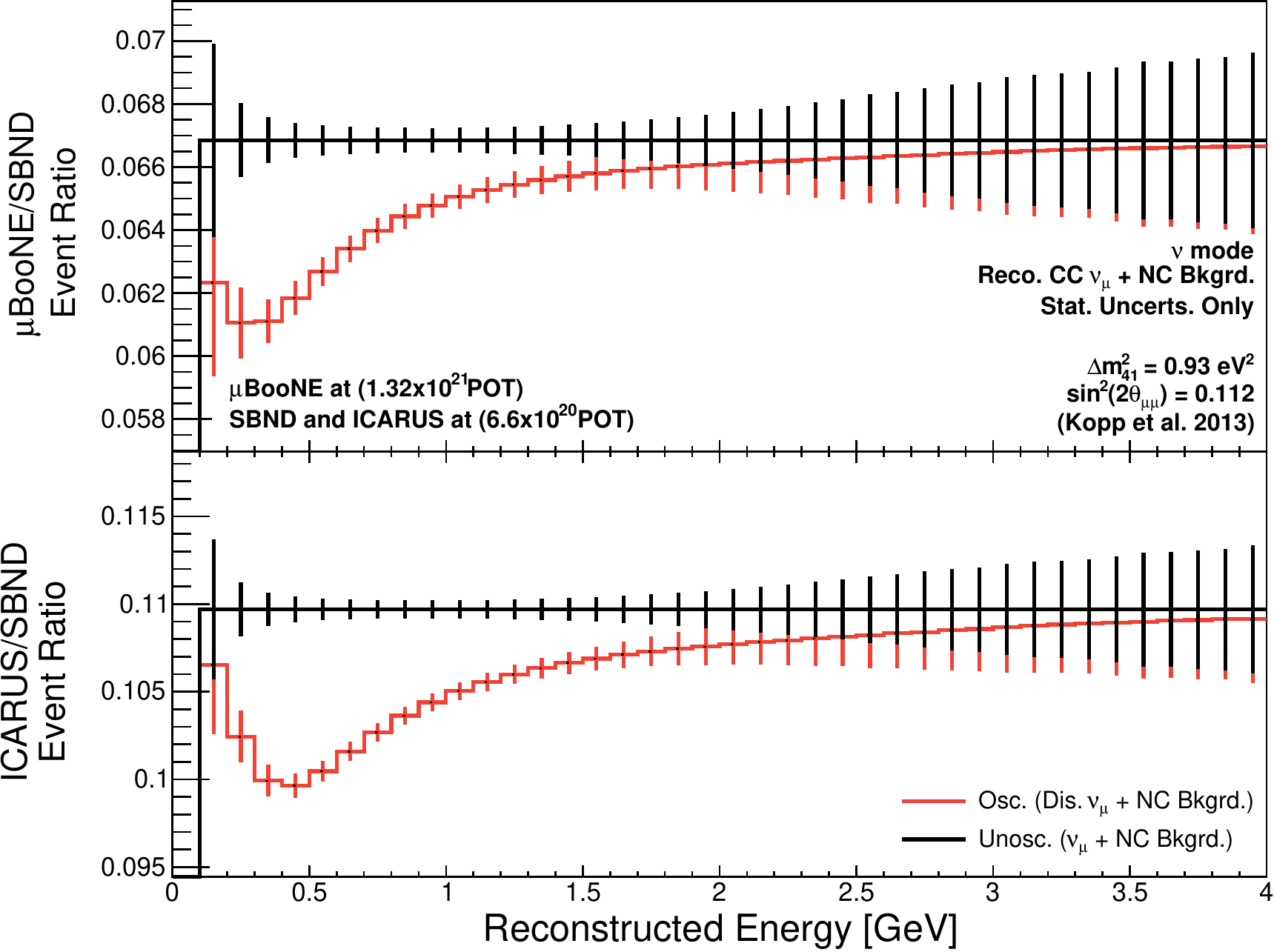}
    \caption{\Pnum CC event ratios as a function of reconstructed neutrino energy showing $\mu$BooNE/SBND (top) and ICARUS/SBND (bottom), applying oscillations generated by Kopp et al. 3+1 best fit parameters. The oscillated (red) and unoscillated (black) cases are shown with statistical uncertainties.}
    \label{fig:mudiseventratio}
\end{figure}

 The energy spectra of SBND, $\mu$BooNE and ICARUS are detailed in Figures \ref{fig:mudissbnd}, \ref{fig:mudismicroboone} and \ref{fig:mudisicarus} respectively, applying global best fit 3+1 parameters from Kopp. 2013\cite{Kopp2013} displayed in Table \ref{tab:bestfits}. Furthermore, the event ratios are shown in Figure \ref{fig:mudiseventratio} and the broad statistics are detailed in Table \ref{tab:mudisstats}.

\begin{table}[h!]
\centering
\caption{\Pnum disappearance channel statistics}
\label{tab:mudisstats}
\begin{tabular}{lp{0.15\linewidth}p{0.15\linewidth}p{0.15\linewidth}} \toprule
 & Tot. Unosc. Events & Tot. Osc. Events & Tot. Deficit \\\midrule
SBND & $5.847\times10^{6}$ & $5.737\times10^{6}$ & $1.099\times10^{5}$ \\ 
$\mu$BooNE & $3.909\times10^{5}$ & $3.716\times10^{5}$ & $1.928\times10^{4}$ \\
ICARUS & $6.414\times10^{5}$ & $6.011\times10^{5}$ & $4.028\times10^{4}$ \\\bottomrule
\end{tabular}
\end{table}

\newpage

\vspace{-0.4cm}
\newpage
\vspace{-0.4cm}
\newpage
For all detectors, the first bin in the spectra (top) is empty due to the energy necessary to create a muon in a CC interaction, hence the first bin of the event ratios (bottom) is ignored. The spectra peaks at a reconstructed energy of \SI{0.95}{\giga\electronvolt} which is of reasonable agreement to Ref. \cite{proposal}. In all plots, the NC misidentification background is plotted but negligible for the \Pnum channel, residing at $\sim$0.041\% of the total signal. For this set of parameters, SBND suffers a 1\% effect in the very low energy bins whereas the far detectors see approximately a 9\% loss at maximum in the far detectors at energies of \SI{0.25}{\giga\electronvolt} and \SI{0.45}{\giga\electronvolt} for $\mu$BooNE and ICARUS respectively.

\subsection{Oscillation Parameter Sensitivities}

The $\Delta m^2_{41} - \sin^22\theta_{\mu\mu}$ representation of observable parameter space for SBN was drawn by creating a 2-dimensional $\chi^2$ surface, with each pixel containing a $\chi^2$ value generated by a particular combination of sterile oscillation parameters. The formulae to obtain this $\chi^2$ are detailed in Section 4.3 and the set of events included are the same as described in Section 5.1.

In order to optimise processing times, the equation for the addition of histograms was arranged as shown in Equation \ref{eq:rearrangemu} such that the range of all $\chi^2$ values, for a given $\Delta m^2_{41}$ value, could be generated by scaling just one set of histograms accordingly by values of $\sin^22\theta_{\mu\mu}$.

\begin{equation}\label{eq:rearrangemu}
N^{\text{Tot.}}_i = \left(\sin^22\theta_{\mu\mu}\right)\cdot N^{\text{Osc, CC} }_{i} + \left(1-\sin^22\theta_{\mu\mu}\right)\cdot N^{\text{Unosc,CC} }_{i} + N^{\text{NC Bkgrd.}}_{i}
\end{equation}

where $N^{\text{Tot.}}_i$ is the energy bin $i$ of the total \Pnum spectra histogram and $N^{\text{Osc, CC}}_i$ corresponds to the CC \Pnum energy bin $i$ with events weighted solely by the oscillating sinusoid of the transition formula as given in Equation \ref{eq:mudisprob}. Also, $N^{\text{Unosc,CC} }_{i}$ and $N^{\text{NC}}_{i}$ correspond to the contents of energy bin $i$ of the unoscillated \Pnum CC and NC background histograms respectively.

\begin{figure}[h!]
    \centering
    \includegraphics[width=0.8\linewidth]{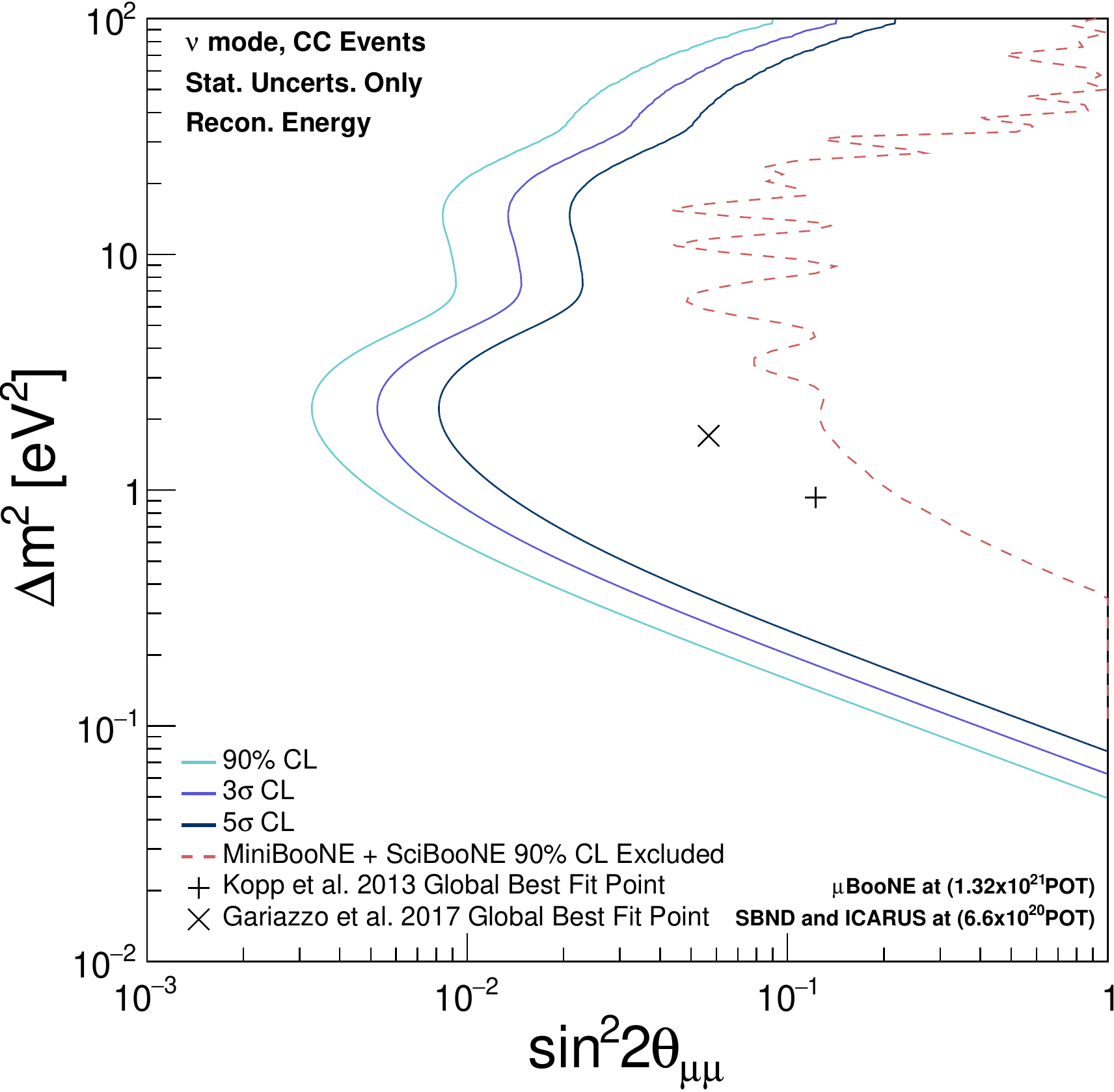}
    \caption{Sensitivity of the combined SBN program to $\Pnum \rightarrow \Pnu_{x}$ oscillations in $\Delta m^2_{41}-\sin^22\theta_{\mu\mu}$ parameter space, with only statistical uncertaintities considered. Frequentist contours corresponding to 90\% (teal), 3$\sigma$ (blue) and 5$\sigma$ (navy) are shown, including space covered by the 90\% limit observed from a combination of MiniBooNE and SciBooNE (red dashed). Also shown are best fit points from two global 3+1 analyses using parameters from Kopp et al. 2013\cite{Kopp2013} (plus) and Gariazzo et al. 2017\cite{2017} (cross).}
    \label{fig:muonmassspace}
\end{figure}

The available sensitivities in $\Delta m^2_{41} - \sin^22\theta_{\mu\mu}$ parameter space are drawn in Figure \ref{fig:muonmassspace}. The resulting plot creates frequentist contours (assuming the Gaussian limit) for 2 degrees of freedom in a 1000 by 1000 $\chi^2$ grid with logarithmic axes.

The best fit points highlighted by 3+1 global analyses Kopp et al.\cite{Kopp2013} and Gariazzo et al.\cite{2017} are well covered by upwards of 5$\sigma$ confidence, with the program's baselines best suited for studying a mass splitting of \SI{1.1}{\electronvolt\squared}. The observed limits from a combined analysis of SciBooNE and MiniBooNE for 90\% CL are also well covered to upwards of 5$\sigma$, providing a further order of magnitude sensitivity to a $\SI{1}{\electronvolt\squared}$ mass splitting in the disappearance amplitude. The sensitivity to mass splittings below \SI{1}{\electronvolt\squared} drops off fast for decreasing sterile masses. This can be interpreted by the SBL oscillation wavelength becoming higher than that of the distance between detectors, thereby rendering the oscillation rate to be increasingly undetectable. The distances between the three SBN detectors can be tracked in this plot, signified by the apparent harmonics in sensitivity for increasing $\Delta m^2_{41}$. Bearing in mind that the two analyses differ in a number of respects, chiefly with the inclusion of systematic uncertainties, there is good qualitative agreement between the contours offered in Figures \ref{fig:muonmassspace} and \ref{fig:sbnmumode}. Both the shapes and available parameter space differ to a minor degree, with Figure \ref{fig:muonmassspace} appearing to be far smoother and covering more space, signified by an approximate 0.06 offset in $\sin^22\theta_{\mu\mu}$. Systematics are therefore considered to be an integral component to an accurate sensitivity prediction.

Using Equation \ref{eq:mu4}, $\sin^22\theta_{\mu\mu}$ was decomposed into a product of functions of the 3+1 mixing angles $\theta_{14}$ and $\theta_{24}$, given by Equation \ref{eq:mumumix}.

\begin{equation}\label{eq:mumumix}
\sin^22\theta_{ee} = 4\left|U_{\mu4}\right|^2\left(1-\left|U_{\mu4}\right|^2\right) = 4\sin^2\theta_{14}\left(1-\sin^2\theta_{14}\right)
\end{equation}

\begin{figure}[h!]
    \centering
    \includegraphics[width=0.8\linewidth]{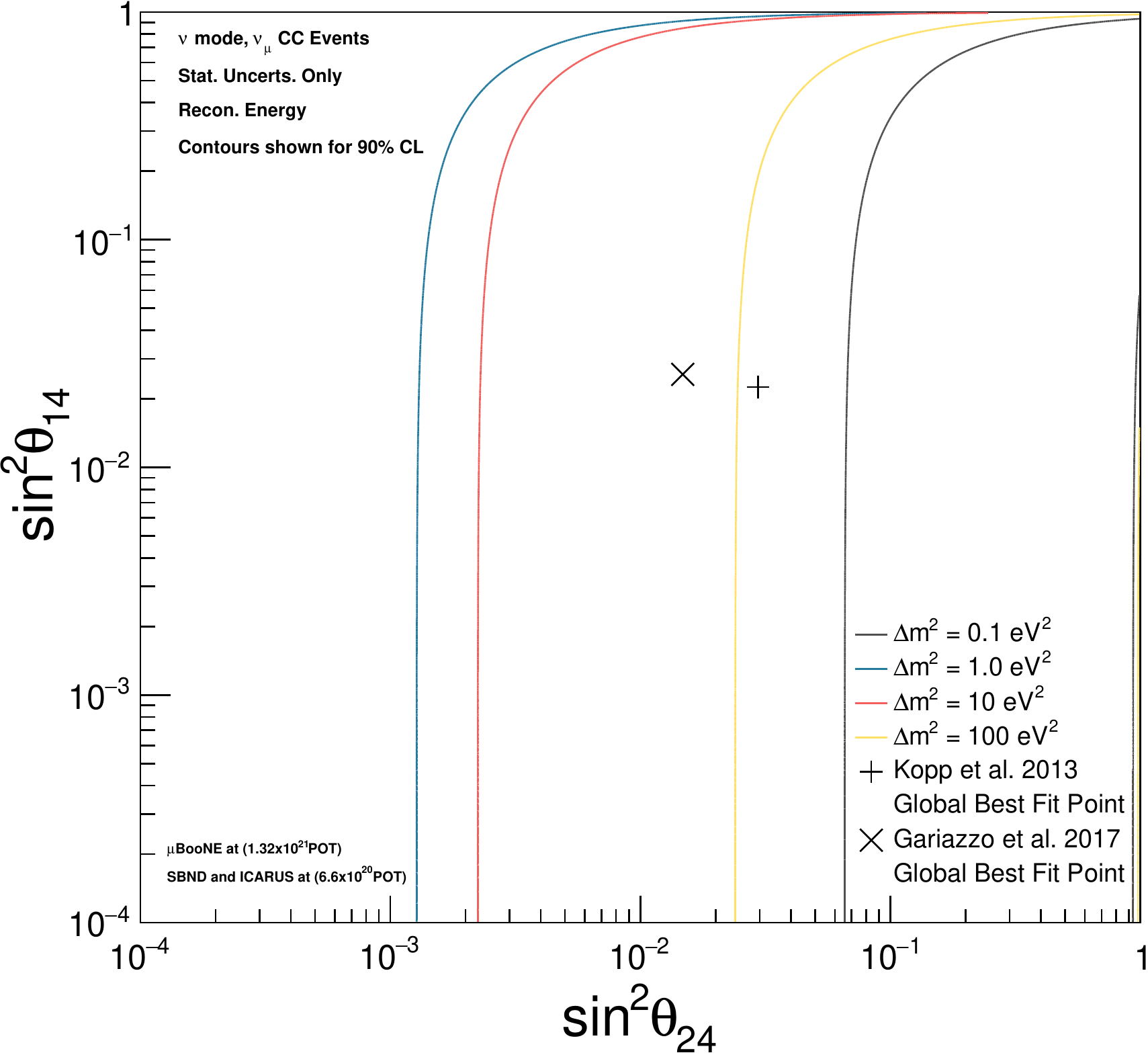}
    \caption{Sensitivity of the combined SBN program to $\Pnum \rightarrow \Pnu_{x}$ oscillations in $\sin^2\theta_{14}-\sin^2\theta_{24}$ parameter space, with only statistical uncertainties considered. Contours of $\Delta m^2_{41}$ set at several different orders of magnitudes are shown at 90\% CL. Also shown are best fit points from two global 3+1 analyses using parameters from Kopp et al. 2013\cite{Kopp2013} (plus) and Gariazzo et al. 2017\cite{2017} (cross). The available parameter space at 90\% CL is the area subtended by the contours on the right hand side.}
    \label{fig:muonanglespace}
\end{figure}

A 1000 by 1000 $\chi^2$ grid was created in the same method as described previously except in $\theta_{14}$ and $\theta_{24}$ mixing angle parameter space, plotting contours for fixed value mass splittings at 90\% CL. The resulting plot is shown in Figure \ref{fig:muonanglespace}.

As reflected in the mixing angle parametrisation of $\sin^22\theta_{\mu\mu}$, as given in Equation \ref{eq:mumumix}, the sensitivity becomes increasing invariant to $\sin^2\theta_{14}$ as it decreases, making the contour dependent mostly on $\sin^2\theta_{24}$ for low mixing between mass states 1 and 4. Also as expected, the sensitivity to $\sin^2\theta_{24}$ is washed away totally as $\sin^2\theta_{14}$ tends to unity. As well, the degenerate nature of mixing angles is reflected by a repeated bunching of contours as $\sin^2\theta_{14}$ tends to unity. Much of the information gained from the previous plot with respect to the mass squared splitting is also evident here. The largest area of parameter space is subtended by the $\Delta m^2_{41} =$ \SI{1}{\electronvolt\squared} contour, furthermore the loss in available sensitivity space from decreasing $\Delta m^2_{41}$ from \SI{1}{\electronvolt\squared} to \SI{0.1}{\electronvolt\squared} is far more drastic than the loss from the increase to \SI{10}{\electronvolt\squared}. As well, only the \SI{1}{\electronvolt\squared} and \SI{10}{\electronvolt\squared} contours contain the two global fit points generously. Therefore, again, the SBN program seems optimised for studying extra mass splittings in the interval from \SIrange[range-phrase = --]{1}{10}{\electronvolt\squared} as highlighted in all global fits. 
\newpage
\section{Electron Appearance Analysis}
The next stage of analysis is to demonstrate the multi-detector sensitivity to SBL \Pnue appearance and disappearance oscillations.
\subsection{Energy Spectra Predictions}

The \Pnue energy spectra incorporated CC events from the intrinsic \Pnue content of the BNB with disappearance weightings applied as given in Equation \ref{eq:edisprob}, in addition to the intrinsic \Pnue NC misidentification background.

\begin{equation}\label{eq:edisprob}
P_{\nu_e \rightarrow \nu_e} = 1 - \sin^2\left(2\theta_{ee}\right)\sin^2\left(\frac{1.27\Delta m^2_{41}L}{E}\right)
\end{equation}

\begin{figure}[h!]
    \centering
    \includegraphics[width=0.705\linewidth]{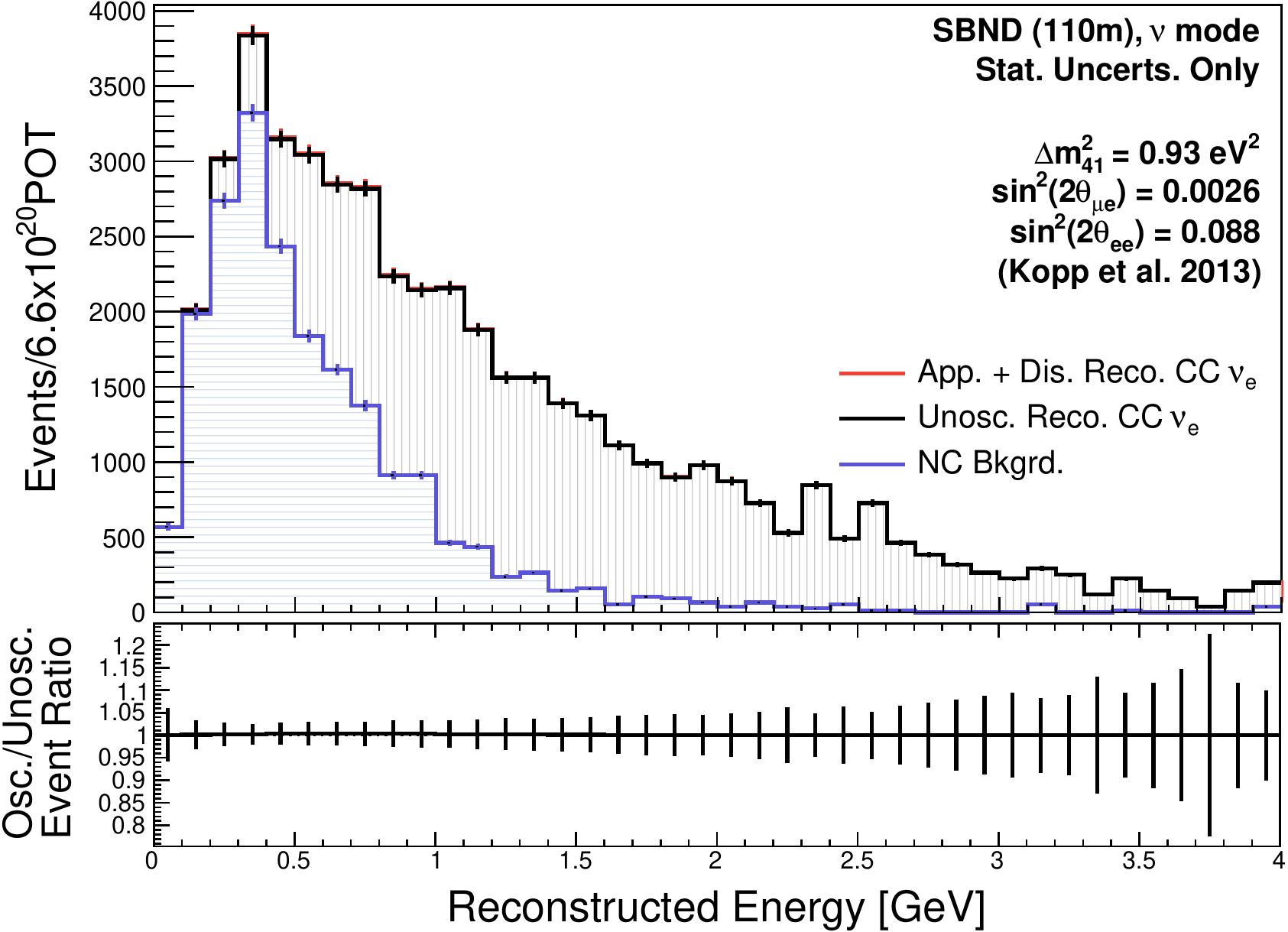}
    \caption{(Top) CC \Pnue event distribution in SBND as a function of reconstructed neutrino energy, applying oscillations generated by Kopp et al. 3+1 best fit parameters. The oscillation part (red), intrinsic beam content (black) and NC background (blue) are presented with statistical uncertainties only. (Bottom) Ratio of oscillated to unoscillated spectra, parallelling the above distribution.}
    \label{fig:electronappsbnd}
\end{figure}

\begin{figure}[h!]
    \centering
    \includegraphics[width=0.705\linewidth]{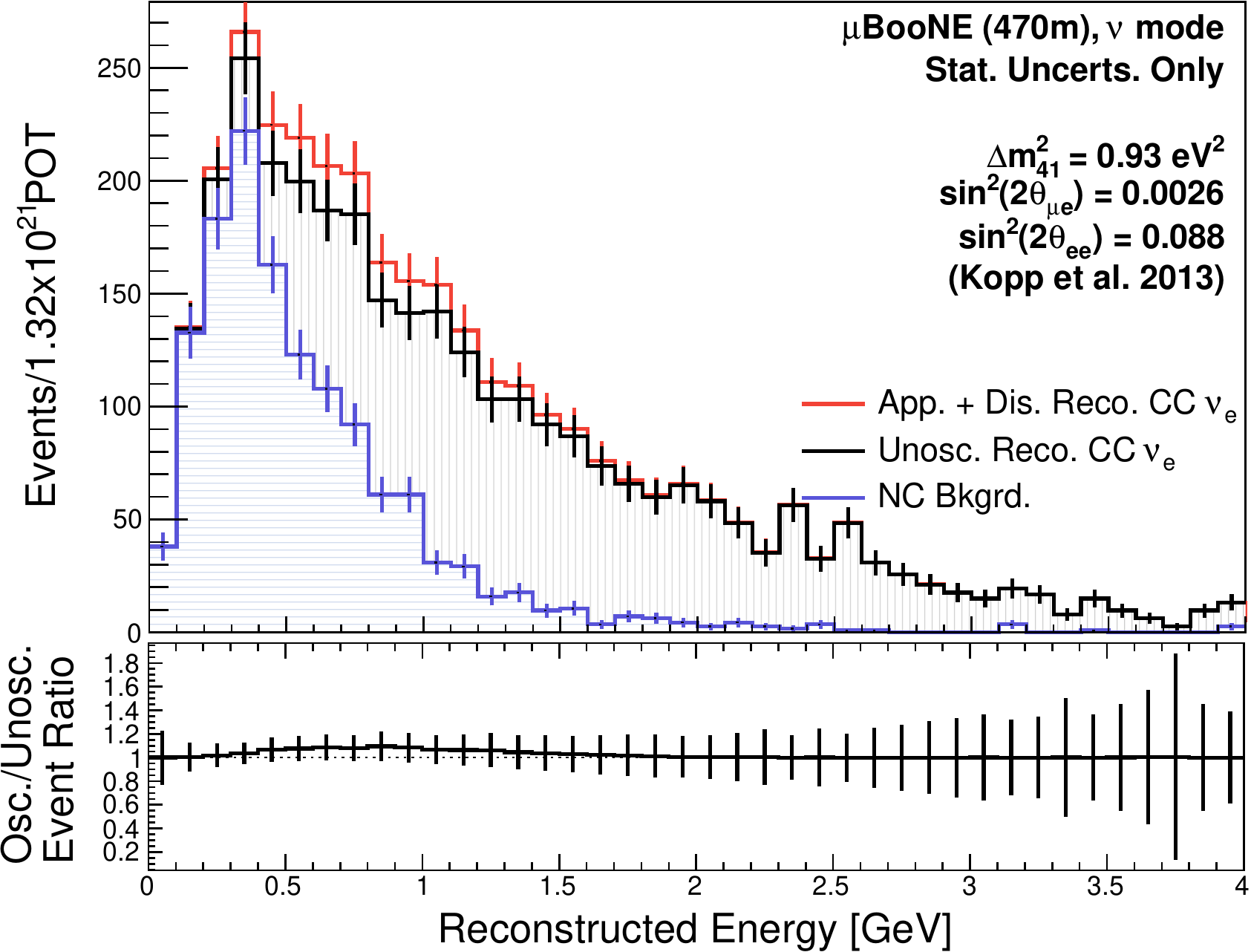}
    \caption{Same as Figure \ref{fig:electronappsbnd} except with event rates observed at $\mu$BooNE.}
    \label{fig:electronappmicroboone}
\end{figure}

\begin{figure}[h!]
    \centering
    \includegraphics[width=0.705\linewidth]{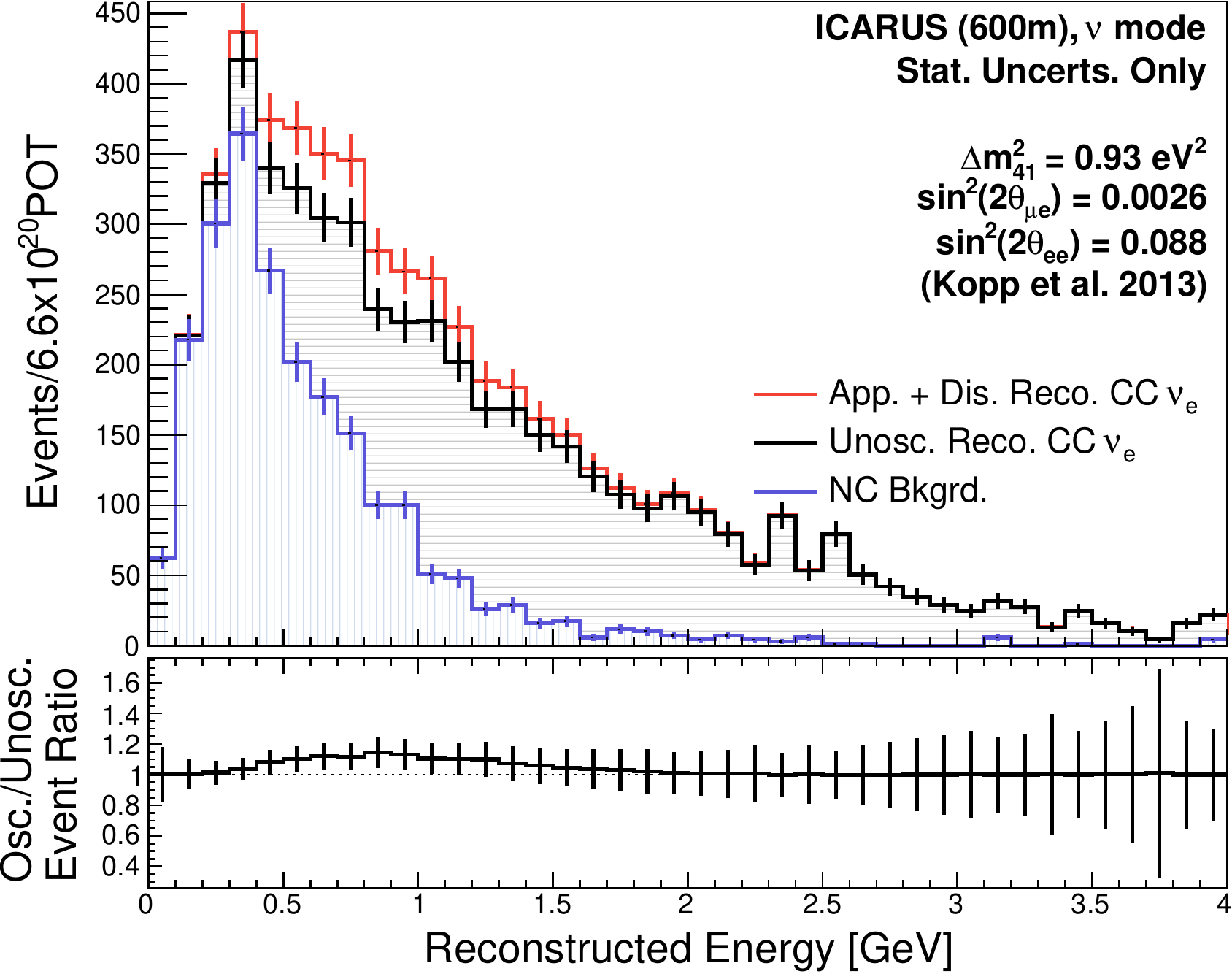}
    \caption{Same as Figure \ref{fig:electronappsbnd} except with event rates observed at ICARUS.}
    \label{fig:electronappicarus}
\end{figure}

\begin{figure}[h!]
    \centering
    \includegraphics[width=0.69\linewidth]{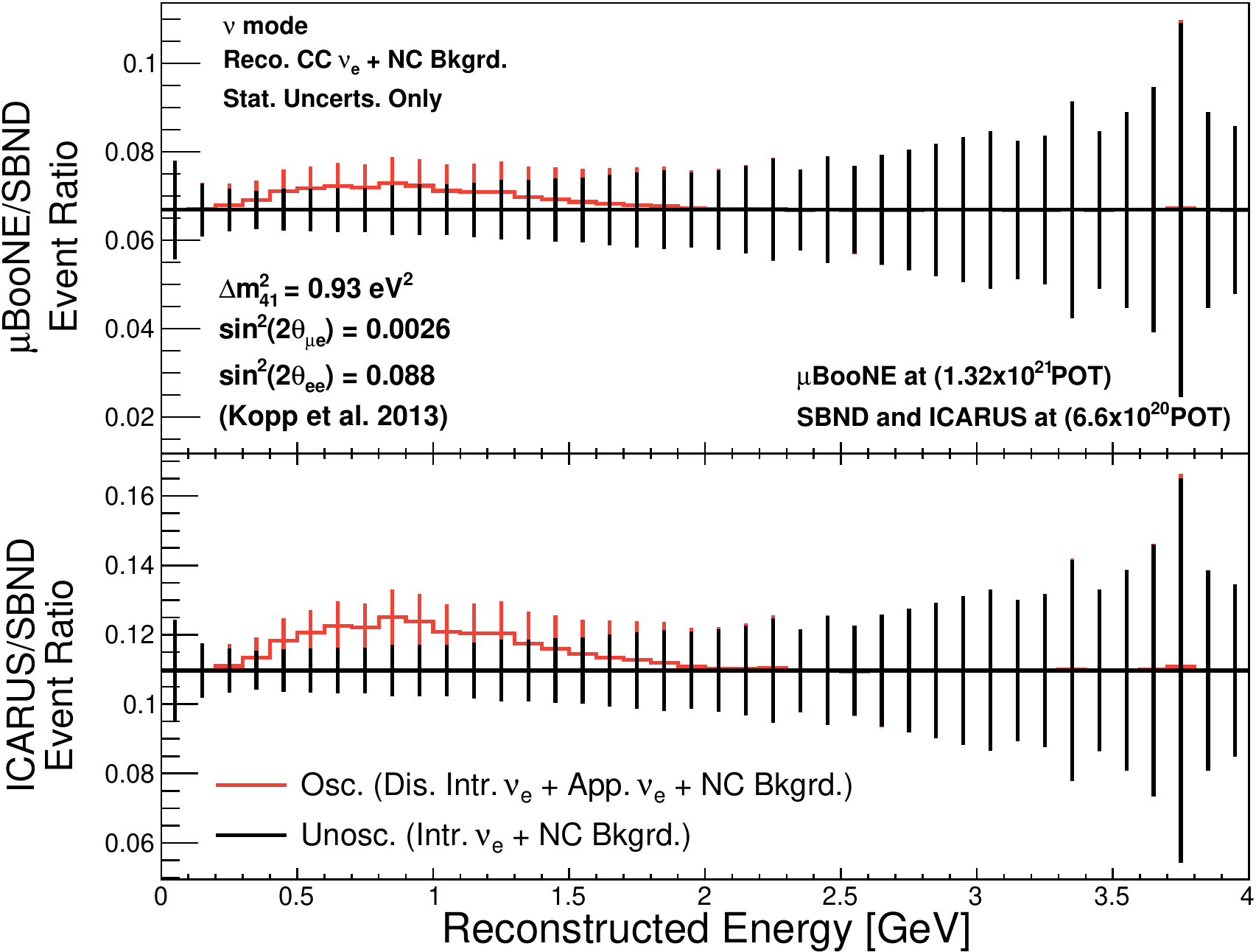}
    \caption{\Pnue CC event ratios as a function of reconstructed neutrino energy showing $\mu$BooNE/SBND (top) and ICARUS/SBND (bottom), applying oscillations generated by Kopp et al. 3+1 best fit parameters. The oscillated (red) and unoscillated (black) cases are shown with statistical uncertainties.}
    \label{fig:electronappeventratio}
\end{figure}

As well, the contribution to CC events resulting from $\Pnum\rightarrow\Pnue$ appearance transitions was also counted. Utilising the inverted data sample containing mostly \Pnue events, the appearance contribution was subject to a weighting as given by Equation \ref{eq:eappprob} and then normalised appropriately by matching the number of NC events in this sample to the number of NC events contained within the intrinsic \Pnue part. This technique was deemed applicable given that NC cross-sections are irrespective of neutrino flavour and also that oscillations of NC events were discarded in this study.

\begin{equation}\label{eq:eappprob}
P_{\nu_\mu \rightarrow \nu_e} = \sin^2\left(2\theta_{\mu e}\right)\sin^2\left(\frac{1.27\Delta m^2_{41}L}{E}\right)
\end{equation}
The impact on 
The energy spectra plots for SBND, $\mu$BooNE and ICARUS are given by Figures \ref{fig:electronappsbnd}, \ref{fig:electronappmicroboone} and \ref{fig:electronappicarus} respectively, with the event ratios and oscillation statistics available in Figure \ref{fig:electronappeventratio} and Table \ref{tab:electronappstats} respectively.

For \Pnue, the appearance channel was predicted to be dominant given the large amount of \Pnum which could transition to \Pnue. The disappearance of \Pnue was thought to be a secondary effect but still however significant when considering of mixing angle sensitivities. The impact of this channel on energy spectra is given in Appendix A.
\begin{table}[h!]
\centering
\caption{\Pnue appearance channel statistics}
\label{tab:electronappstats}
\begin{tabular}{lp{0.15\linewidth}p{0.15\linewidth}p{0.15\linewidth}} \toprule
 & Tot. Unosc. Events & Tot. Osc. Events & Tot. Excess \\\midrule
SBND &$4.740\times10^{4}$ & $4.750\times10^{4}$ & $9.754\times10^{1}$  \\ 
$\mu$BooNE & $3.169\times10^{3}$ & $3.308\times10^{3}$ & $1.398\times10^{2}$ \\
ICARUS & $5.199\times10^{3}$ & $5.526\times10^{3}$ & $3.260\times10^{2}$ \\ \bottomrule
\end{tabular}
\end{table}

\newpage

\newpage
\newpage
The NC misidentification background especially dominates the sub-\SI{0.7}{\giga\electronvolt} bins in the \Pnue sample, which is unexpected given the efficiencies applied in Section 4.1. Generally, the uncertainties are significantly larger due to the drop in statistics as compared to the \Pnum sample. In SBND, the appearance signal is negligible with respect to statistical uncertainties gained from a \SI{6.6e20}{\POT} beam exposure, which renders it useful for characterising the beam. In the far detectors, the significance of the appearance signal corresponds to a 14\% and 18\% excess at maximum in $\mu$BooNE and ICARUS respectively.


\subsection{Oscillation Parameter Sensitivities}

With the same method as described in Section 6.2 for the \Pnum data, a 1000 by 1000 $\chi^2$ surface was drawn except using \Pnue appearance spectra to generate sensitivities to the appearance channel amplitude $\sin^22\theta_{\mu e}$, which is the dominant oscillation mode in the \Pnue channel. The equation for the contents of energy bin $i$ in the histogram of total events for fixed $\Delta m^2_{41}$ is shown in Equation \ref{eq:eappform}, in which the disappearance signal is discarded.

\begin{equation} \label{eq:eappform}
N_i^{\text{Tot.}} = N^{\text{Intr., CC}}_{i} + N^{\text{Intr., NC Bkgrd.}}_{i} + \left(\sin^22\theta_{\mu e}\right)\cdot N^{\text{App., CC}}_i
\end{equation}

where $N^{\text{Intr., CC}}_{i}$ and $N^{\text{Intr., NC Bkgrd.}}_{i}$ are the contents of energy bin $i$ for the unoscillated intrinsic CC \Pnue content and NC misidentification background respectively. $N^{\text{App., CC}}_i$ is the contents of bin $i$ in the energy spectra of the normalised appearance signal.

\begin{figure}[h!]
    \centering
    \includegraphics[width=0.8\linewidth]{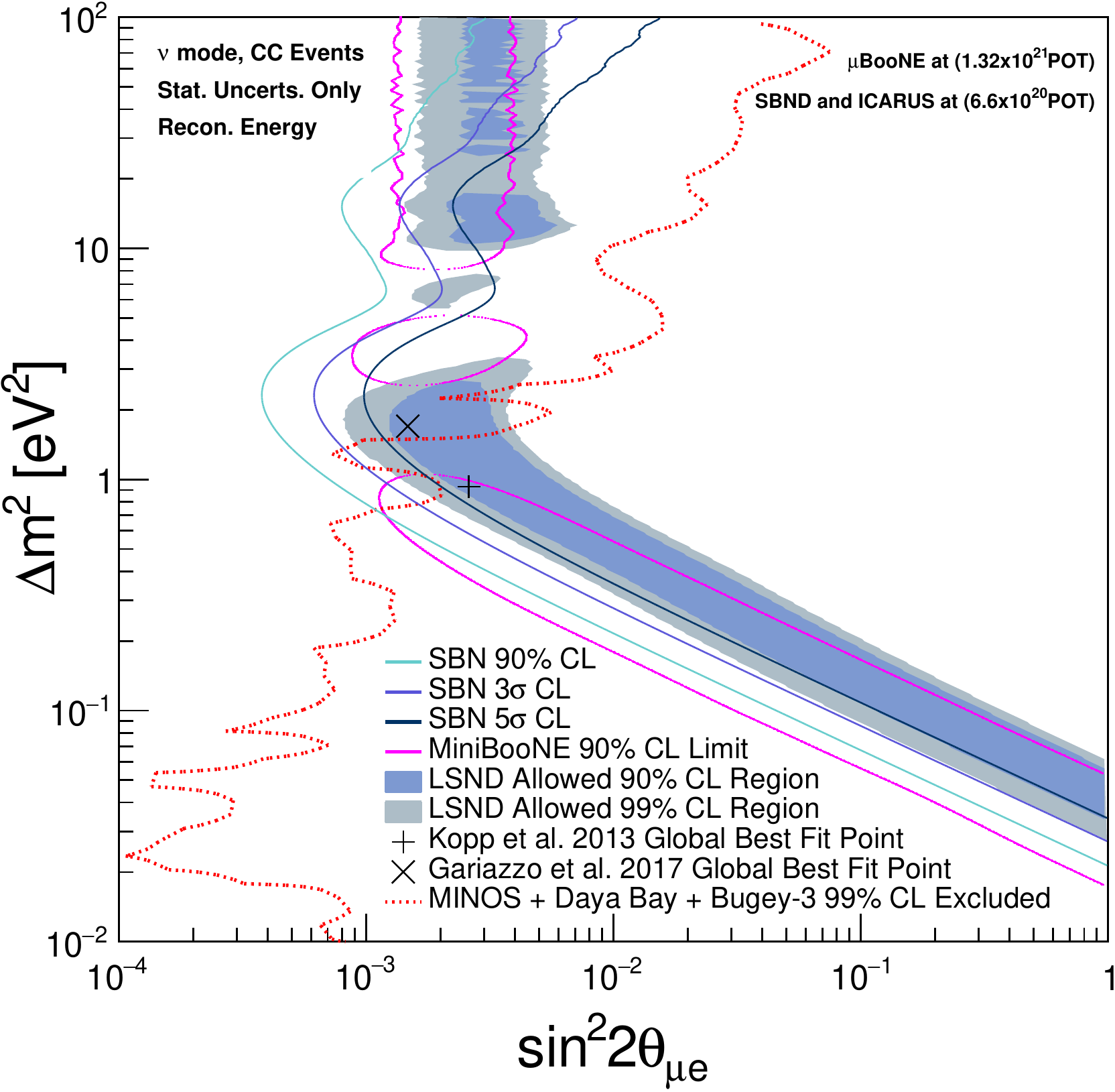}
    \caption{Sensitivity of the combined SBN program to $\Pnum \rightarrow \Pnu_{e}$ oscillations in $\Delta m^2_{41}-\sin^22\theta_{\mu e}$ parameter space, with only statistical uncertaintities considered. Frequentist contours corresponding to 90\% (teal), 3$\sigma$ (blue) and 5$\sigma$ (navy) are shown, also including the 90\% observed limit from MiniBooNE\cite{miniboone} (pink), 99\% excluded region from a combined analysis of MINOS, Daya Bay and Bugey-3\cite{minos} (red dashed) and the LSND allowed parameter region\cite{lsnd} at both 90\% (steel blue) and 99\% CL (grey). Also shown are best fit points from two global 3+1 analyses using parameters from Kopp et al. 2013\cite{Kopp2013} (plus) and Gariazzo et al. 2017\cite{2017} (cross).}
    \label{fig:electronmassspace}
\end{figure}

Figure \ref{fig:electronmassspace} shows the predicted sensitivities of SBN in $\Delta m^2_{41}-\sin^22\theta_{\mu e}$ parameter space, displaying frequentist contours in addition to various experimental data for reference.

In comparison to the \Pnum sensitivities, sensitivity to $\sin^22\theta_{\mu e}$ at maximum covers a further order of magnitude even at 5$\sigma$. The best fit points of global analyses are also well covered by upwards of $5\sigma$ confidence. As expected, the \Pnue appearance plot appears as an offset version of the \Pnum disappearance plot (Figure \ref{fig:muonmassspace}) though with seemingly less of a sensitivity drop off for high values of $\Delta m^2_{41}$. The 99\% LSND region for \SI{1}{\electronvolt\squared} is well covered at $5\sigma$, therefore a conclusive refutation of the LSND anomaly should be possible given the beam exposures applied, statistically speaking. The region defined by the MiniBooNE 90\% CL limit is covered by SBN to a lesser degree, therefore decisive studies of the MiniBooNE anomaly may however require a combined run analysis. Nevertheless, from a statistical point of view, it appears SBN provides a world leading sensitivity to oscillations generated from \SI{1}{\electronvolt\squared} and \SI{10}{\electronvolt\squared} mass splittings. Also evident here is the mentioned tension between MiniBooNE and LSND, since these highlighted parameter regions only partially overlap. Furthermore, exclusion regions from a combined view of MINOS, Daya Bay and Bugey-3 sterile searches seem to rule out large areas of potential regions highlighted by MiniBooNE and LSND especially for sub-eV mass splittings.

The disappearance channel was enabled again when drawing sensitivities in mixing angle space for fixed values of $\Delta m^2_{41}$. The equation for the number of events in energy bin $i$ of the the total \Pnue CC spectra histogram is shown in Equation \ref{eq:edisform}.

\begin{multline} \label{eq:edisform}
N_i^{\text{Tot.}} = \left(\sin^22\theta_{ee}\right)\cdot N^{\text{Osc. Intr. CC}}_{i} + \left(1 - \sin^22\theta_{ee}\right)\cdot N^{\text{Unosc. Intr. CC}}_{i} \\+ N^{\text{Intr. NC}}_{i} + \left(\sin^22\theta_{\mu e}\right)\cdot N^{\text{App, CC}}_i
\end{multline}

Using Equations \ref{eq:e4} and \ref{eq:mu4}, channel amplitudes $\sin^22\theta_{\mu e}$ and $\sin^22\theta_{ee}$ were decomposed into their representations as a function of mixing angles $\theta_{14}$ and $\theta_{24}$ as given in Equations \ref{eq:muemix} and \ref{eq:eemix}.

\begin{equation}\label{eq:muemix}
\sin^22\theta_{\mu e} = 4\left|U_{\mu4}\right|^2\left|U_{e4}\right|^2 = 4\left(1-\sin^2\theta_{14}\right)\sin^2\theta_{24}\sin^2\theta_{14}
\end{equation}

\begin{equation}\label{eq:eemix}
\sin^22\theta_{ee} = 4\left|U_{e4}\right|^2\left(1-\left|U_{e4}\right|^2\right) = 4\sin^2\theta_{14}\left(1-\sin^2\theta_{14}\right)
\end{equation}

where $N^{\text{Osc. Intr. CC}}_{i}$ represents the contents of energy bin $i$ for the intrinsic CC \Pnue flux contribution weighted by the oscillating sinusoid in Equation \ref{eq:edisprob} and $N^{\text{Unosc. Intr. CC}}_{i}$ is the contents of energy bin $i$ for the unoscillated intrinsic \Pnue flux spectra.

\begin{figure}[h!]
    \centering
    \includegraphics[width=0.8\linewidth]{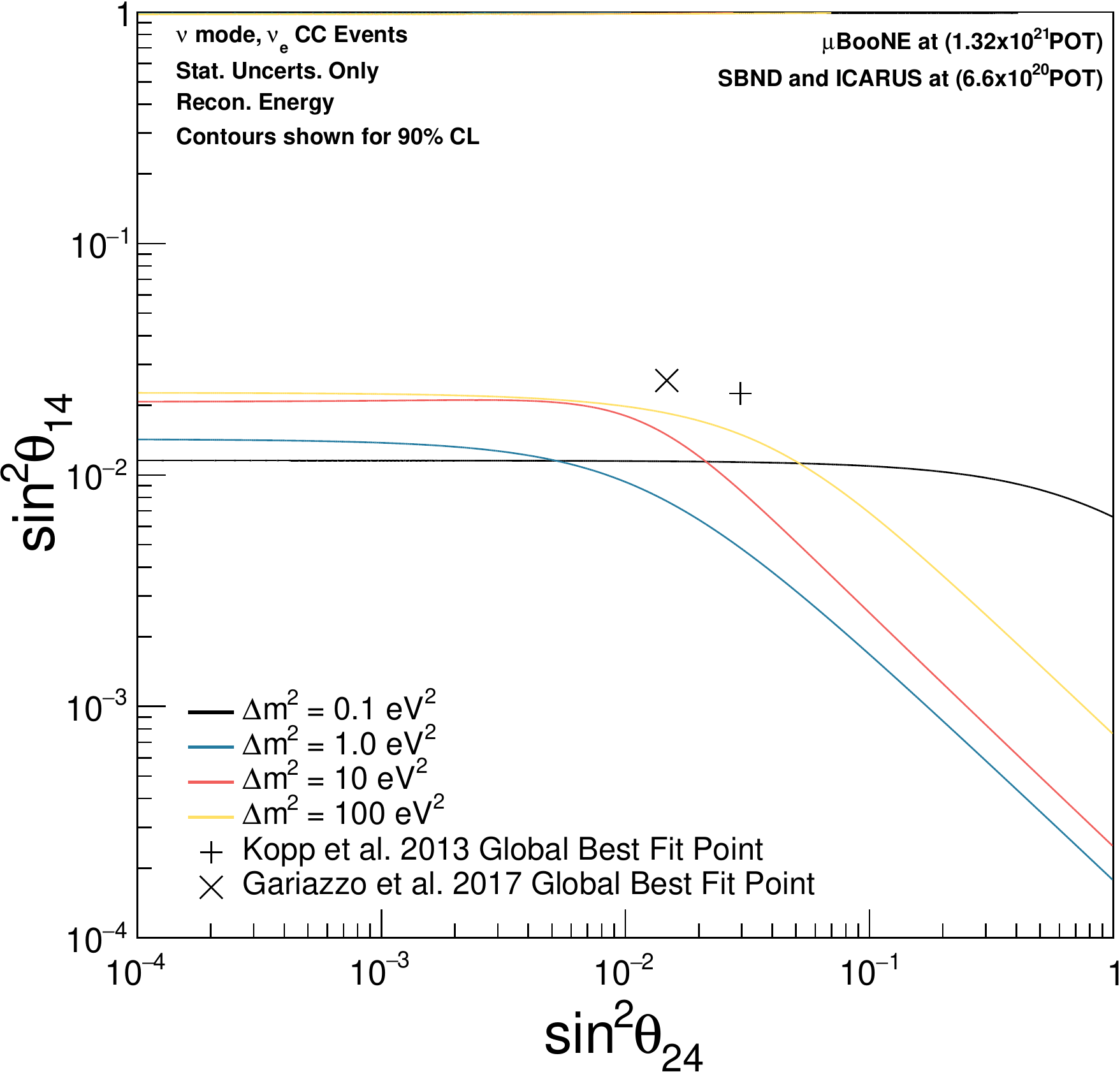}
    \caption{Sensitivity of the combined SBN program to $\Pnum \rightarrow \Pnu_{e}$ oscillations in $\sin^2\theta_{14}-\sin^2\theta_{24}$ parameter space, with only statistical uncertainties considered. Contours of $\Delta m^2_{41}$ set at several different orders of magnitudes are shown at 90\% CL. Also shown are best fit points from two global 3+1 analyses using parameters from Kopp et al. 2013\cite{Kopp2013} (plus) and Gariazzo et al. 2017\cite{2017} (cross).}
    \label{fig:electronappparamspace}
\end{figure}

The sensitivity contours in mixing angle parameter space plotted for different orders of magnitude of $\Delta m^2_{41}$ at 90\% CL are presented in Figure \ref{fig:electronappparamspace} with two global best fit points for reference.

The appearance contribution to signal is evident from the contour deflection for high values of $\sin^2\theta_{24}$, since appearance depends on both $\theta_{14}$ and $\theta_{24}$. The disappearance signal creates a base sensitivity space which depends weakly on the mass splitting, displayed by the residual sensitivity for low values of $\sin^2\theta_{24}$. While the available parameter space is largest for mass splittings of \SI{1}{\electronvolt\squared}, it is interesting to note that the case of a mass splitting of \SI{0.1}{\electronvolt\squared} creates a marginally greater sensitivity to $\theta_{14}$. The global fit points are covered at 90\% by the full range of mass squared splittings.
\newpage
\section{Combined Channel Analysis}

\begin{figure}[h!]
    \centering
    \includegraphics[width=0.8\linewidth]{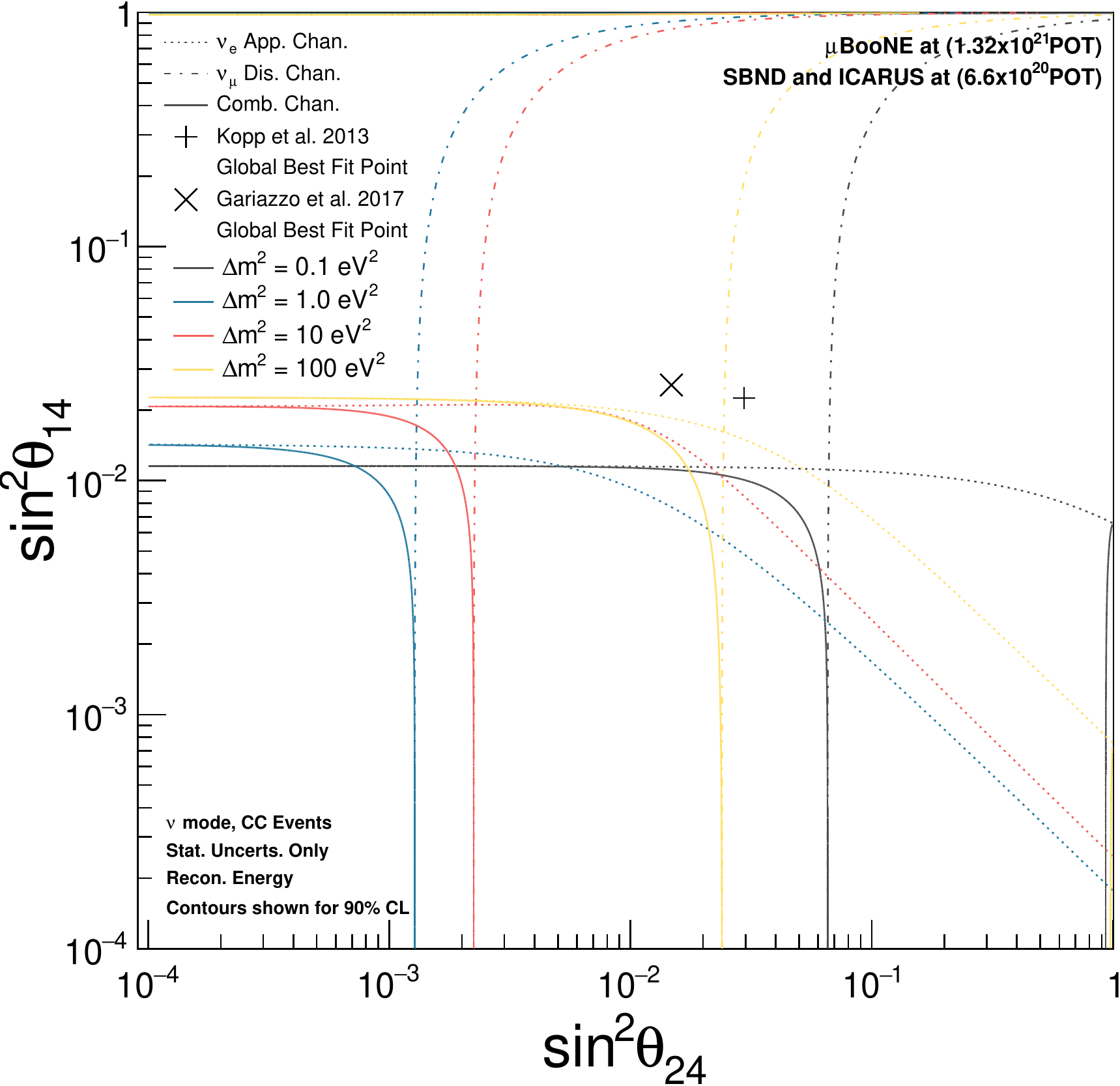}
    \caption{Combined sensitivities of the SBN program in $\sin^2\theta_{14}-\sin^2\theta_{24}$ parameter space achievable through analysis of $\Pnum \rightarrow \Pnu_x$ (dashed and dotted) and $\Pnum \rightarrow \Pnue$ (dotted) channels. Contours of $\Delta m^2_{41}$ set at several different orders of magnitudes are shown at 90\% CL with only statistical uncertainties considered. Also shown are best fit points from two global 3+1 analyses using parameters from Kopp et al. 2013\cite{Kopp2013} (plus) and Gariazzo et al. 2017\cite{2017} (cross).} 
    \label{fig:combined}
\end{figure}

The two $\chi^2$ surfaces obtained from the \Pnum disappearance and \Pnue appearance channels in mixing angle space given by Figures \ref{fig:muonanglespace} and \ref{fig:electronappparamspace} respectively, were simply added together in order to form a combined sensitivity representation. In Figure \ref{fig:combined}, contours for 90\% CL are plotted showing the combined sensitivities and also the two constituent channels individually.

The combination of channels creates a remarkably uniform and tightly defined region of parameter space observable to 90\% confidence. Seemingly, analysis of the \Pnue appearance channel is best suited for determining $\theta_{14}$ whereas analysis of the \Pnum disappearance channel is best for determining $\theta_{24}$. The new area in parameter space observable gained by a combined channel analysis corresponds to the small area subtended by the combined contour and individual channel contours. Nevertheless, the full sensitivity of the SBN program to the range of mixing angles $\theta_{14}$ and $\theta_{24}$ is evident. The \Pnue and \Pnum event samples are highly complementary and together provide strong sensitivities to the main parameters of the 3+1 model.

\section{Conclusions and Further Work}

It has been shown, as a world's first, the available sensitivities of the SBN program to SBL light sterile neutrino oscillations from combining \Pnue appearance and \Pnum disappearance channels. In determining the SBL oscillation parameters, it has been shown that both \Pnue appearance and \Pnum disappearance channels are as essential as each other when it comes to providing high coverage of values of $\theta_{14}$ and $\theta_{24}$. Assuming the given beam exposures and statistical uncertainties only, SBN should be able to provide full $5\sigma$ coverage of the LSND 99\% observed space through the \Pnue appearance channel in addition to all global fit best points in both channels. SBN will also be able to search for SBL \Pnum disappearance at amplitudes a further order of magnitude than that observed by MiniBooNE and SciBooNE to high confidence, statistically speaking. Overall, the sensitivities in parameter space with respect to the mass splitting and mixing angle parameters are in satisfactory agreement to those already produced in the SBN proposal paper\cite{proposal}.

Though battling statistical limitations in neutrino experiments is always a major issue, solely considering statistical uncertainties is merely a first step when it comes to producing sensitivities which are more representative of the experiment in action. The process of including the range of systematic uncertainties would prove to be a demanding but intriguing study of it's own and so is left for future contributors to embark on. The sources of these systematics may include the beam flux, interaction cross-sections, reconstructed energy and normalisations of signal and background. The full range of background signal contributions needs to be addressed also in the future. For example, the `dirt' events produced from interactions outside of the detector material as well as cosmogenics are both considered to be prominent backgrounds. To consider more specifically the true event types of NC misidentifications may also be an insightful study. Furthermore, an optimal variable binning scheme for energy spectra plots should be employed in the future. Additionally, although a 2$\nu$ approximated oscillation model should be sufficient in describing SBL oscillations, a full 4$\nu$ analysis should be adopted in future sensitivity studies to allow for precise predictions. Finally, investigating the significance of the oscillation of NC interactions in a 3+1 model would also be an intriguing future study to undertake.

The detection of new fundamental particles at SBN would be a groundbreaking discovery, increasing the current disparity between SM neutrinos and experimentation further still. Being a relatively young field, the next generation of oscillation experiments leaves a wide range of opportunities for the discovery of new physics, bringing about a true golden age for neutrino physics.

\vspace{0.5cm}

\PRLsep
\fancyhead[LE,RO]{}

\newpage
\appendix
\section{Significance of the Electron Disappearance Channel}

\begin{figure}[h!]
    \centering
    \includegraphics[width=0.9\linewidth]{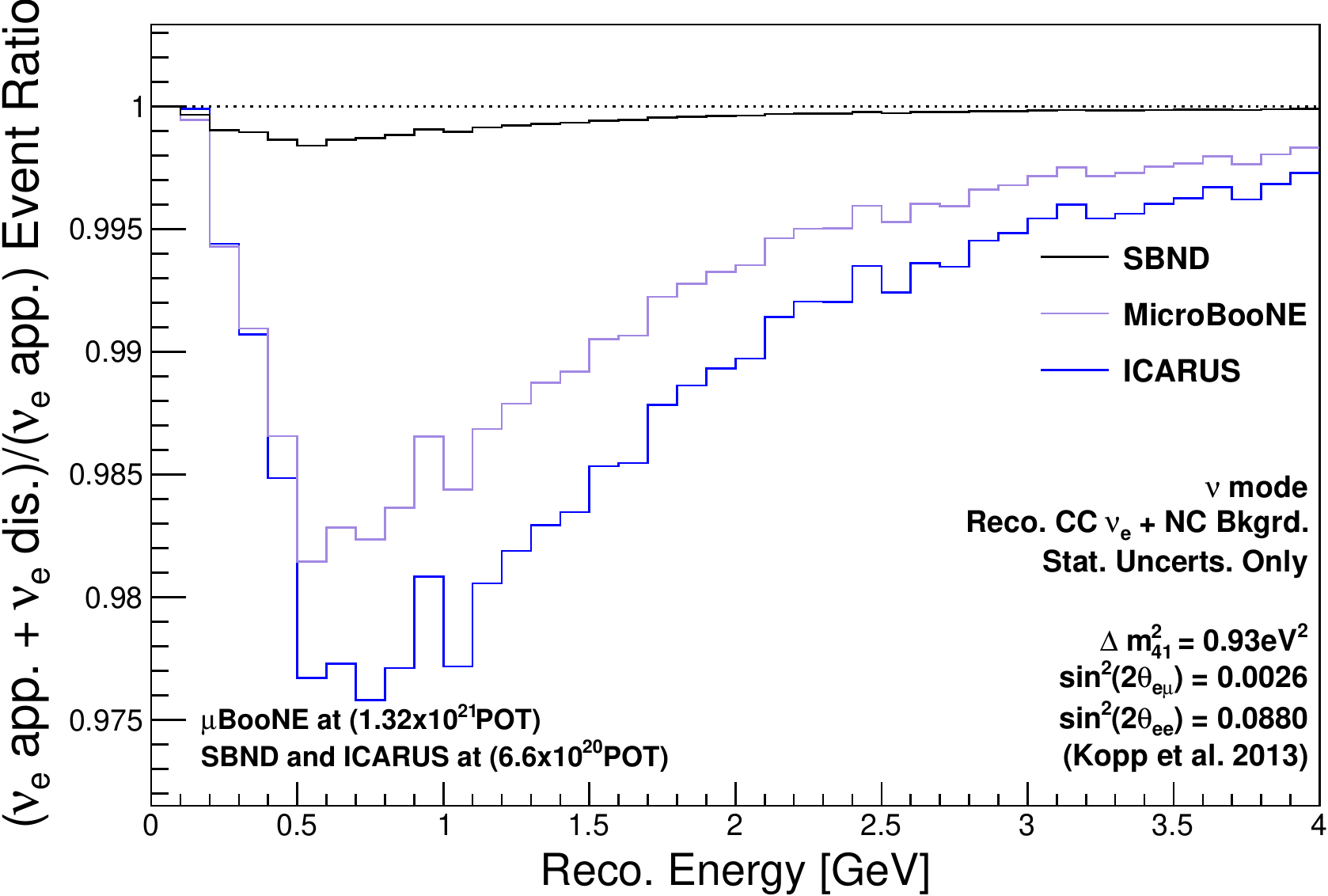}
    \caption{Significance of the disappearance channel on CC \Pnue energy spectra for each SBN experiment applying Kopp et al. 2013\cite{Kopp2013} global best fit parameters.}
    \label{fig:electronappdiseffect}
\end{figure}

In the far detectors, disappearance creates an approximate 2\% loss in \Pnue events whereas in SBND, the effect is much less significant for these set of parameters. The disappearance signal is however deemed an essential effect when it comes to drawing sensitivities to the SBL mixing angles not only due to the potential impact on spectra. The disappearance channel amplitude depends only on $\theta_{14}$ in the SBL approximation whereas the appearance channel depends on both $\theta_{14}$ and $\theta_{24}$. Therefore, studies of \Pnue disappearance effects should create a residual base sensitivity as $\theta_{24}$ tends to zero.
\newpage
\section{Effect on Sensitivities of the Decay Point Distribution}
\begin{figure}[h!]
    \centering
    \includegraphics[width=0.9\linewidth]{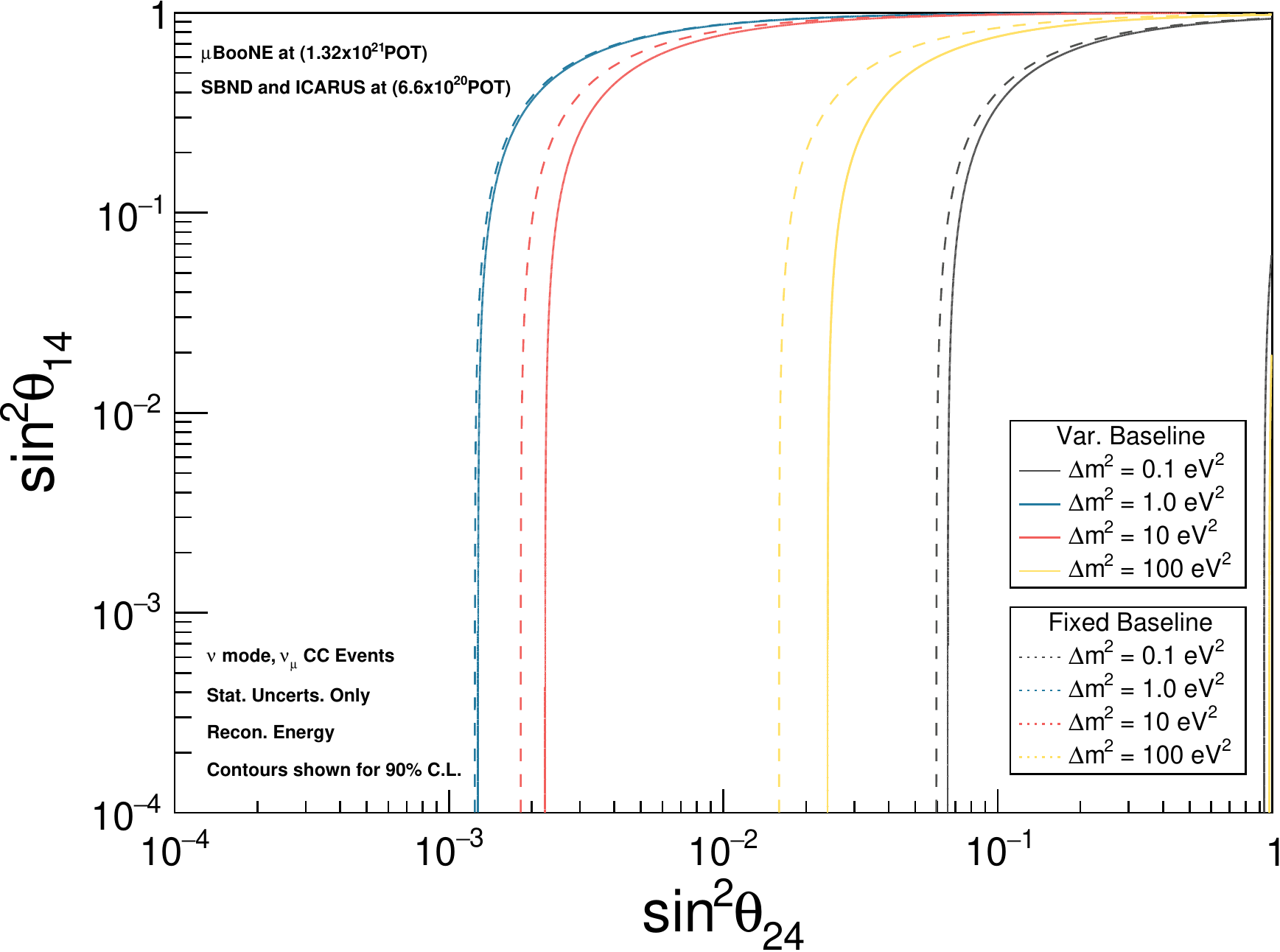}
    \caption{Difference in sensitivities of the combined SBN program to $\Pnum \rightarrow \Pnu_{x}$ oscillations in $\sin^2\theta_{14}-\sin^2\theta_{24}$ parameter space for variable (solid) and fixed baseline (dashed) algorithms, with only statistical uncertainties considered. Contours of $\Delta m^2_{41}$ set at several different orders of magnitudes are shown at 90\% CL}
    \label{fig:muonanglediff}
\end{figure}

The mixing angle sensitivity plots in Figures \ref{fig:muonanglespace} and \ref{fig:electronappparamspace} were drawn again with fixed baselines in order to study the drop in sensitivities due to the decay distribution effect. The baseline applied to the oscillation weightings in this case was set to the fixed distance between the respective detector and the BNB target, as given in Table 3. Figure \ref{fig:muonanglediff} shows the reduction in mixing angle sensitivities in the \Pnum disappearance channel from when variable baselines are considered as opposed to fixed baselines.

\begin{figure}[h!]
    \centering
    \includegraphics[width=0.9\linewidth]{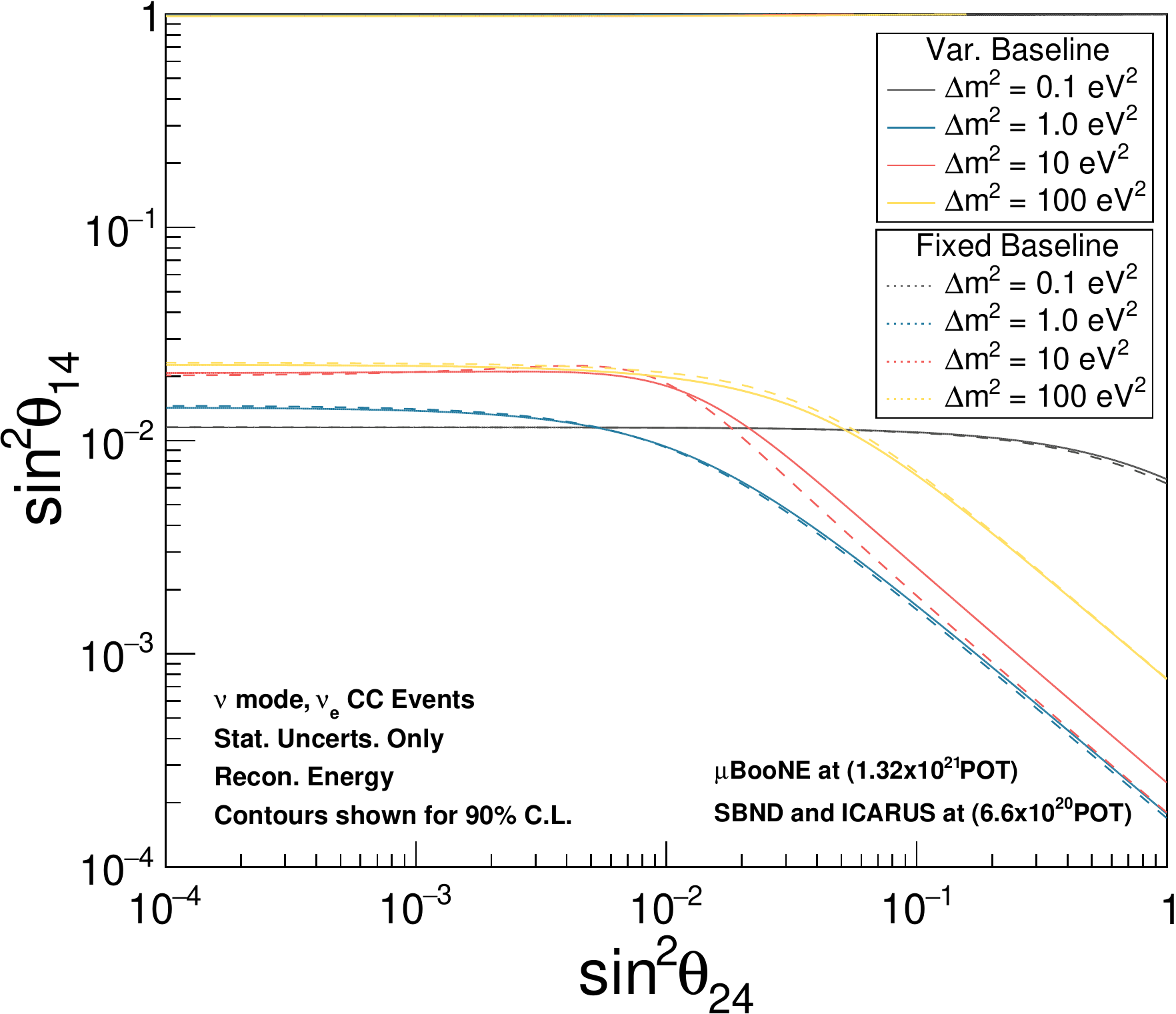}
    \caption{Difference in sensitivities of the combined SBN program to $\Pnum \rightarrow \Pnue$ oscillations in $\sin^2\theta_{14}-\sin^2\theta_{24}$ parameter space for variable (solid) and fixed baseline (dashed) algorithms, with only statistical uncertainties considered. Contours of $\Delta m^2_{41}$ set at several different orders of magnitudes are shown at 90\% CL}
    \label{fig:electronappdiff}
\end{figure}

The loss in space is most evident for $\Delta m^2_{41}$ values of \SI{10}{\electronvolt\squared} and \SI{100}{\electronvolt}. This is expected since for higher values of $\Delta m^2_{41}$ the shape of oscillations becomes more important in analysis, a variable baseline smears this shape away and thus reduces sensitivities.
Figure \ref{fig:electronappdiff} shows the same as in Figure \ref{fig:muonanglediff} except with mixing angle sensitivities available from a \Pnue appearance analysis.
\newpage

The loss in space in the \Pnue channel is mostly negligible apart from the loss generated by the \SI{10}{\electronvolt\squared} contour, of which the cause is not understood. In summary, accounting for the distribution of decay points in the decay pipe in future precision sensitivity studies should be a necessary endeavour.


\begin{thebibliography}{99}
\bibitem{mass} N. Palanque-Delabrouille et al. (2015) `Constraint on neutrino masses from SDSS-III/BOSS Ly$\alpha\alpha$ forest and other cosmological probes', JCAP 1502 2\textbf{(2015)}. doi:10.1088/1475-7516/2015/02/045.
\bibitem{superk} K. Abe et al. \textit{The T2K Collaboration} (2012) `First muon-neutrino disappearance study with an off-axis beam', Phys. Rev. D, 3\textbf{(85)} 031103(R). doi:10.1103/PhysRevD.85.031103.
\bibitem{sno1} B. Aharmim et al. \textit{SNO Collaboration} (2013) `Combined analysis of all three phases of solar neutrino data from the Sudbury Neutrino Observatory', Phys. Rev. C, 2\textbf{(88)} 025501. doi:10.1103/PhysRevC.88.025501.
\bibitem{portecorvo} B. Pontecorvo (1957) `Mesonium and anti-mesonium', Sov. Phys. JETP, \textbf{(6)}, 429-429.
\bibitem{portecorvo1} B. Pontecorvo (1967) `Neutrino experiments and the problem of conservation of leptonic charge', Sov. Phys. JETP, 26\textbf{(5)}, 984-988.
\bibitem{pdg} C. Patrignani et al. \textit{Particle Data Group} (2016) `Review of Particle Physics', Chin. Phys. C, \textbf{(40)} 100001. doi:10.1088/1674-1137/38/9/090001.
\bibitem{nova} P. Adamson et al. \textit{NOvA Collaboration} (2016) `First Measurement of Electron Neutrino Appearance in NOvA', Phys. Rev. Lett., 15\textbf{(116)} 151806. doi:10.1103/PhysRevLett.116.151806.
\bibitem{dune} R. Acciarri \textit{The DUNE collaboration} (2016) `Long-Baseline Neutrino Facility (LBNF) and Deep Underground Neutrino Experiment (DUNE) : Volume 1: The LBNF and DUNE Projects', FERMILAB-DESIGN-2016-01. arXiv:1601.05471.
\bibitem{t2k} K. Abe et al. \textit{T2K Collaboration} (2017) `Combined Analysis of Neutrino and Antineutrino Oscillations at T2K', Phys. Rev. Lett., 15\textbf{(118)} 151801. doi:10.1103/PhysRevLett.118.151801.
\bibitem{hyperk} K. Abe et al. \textit{Hyper-Kamiokande Working Group} (2015) `Physics potential of a long baseline neutrino oscillation experiment using J-PARC neutrino beam and Hyper-Kamiokande', Prog. Theor. Exp. Phys., 5\textbf{(2015)} 053C02.
\bibitem{n0bb} S. Dell'Oro et al. (2016) `Neutrinoless Double Beta Decay: 2015 Review', Adv. in High Ene. Phys, \textbf{(2016)} 2162659. doi:10.1155/2016/2162659
\bibitem{sno} S. Andringa et al. \textit{The SNO+ Collaboration} (2016) `Current status and future prospects of the SNO+ experiment', Adv. in High Ener. Phys, \textbf{(2016)} 6194250. doi:10.1155/2016/6194250.
\bibitem{supernemo} H. Gómez (2016) `Recent results from the NEMO-3 experiment and the status of SuperNEMO', Nuc. and Part. Phys. Proc., \textbf{(273–275)}, 1765-1770. doi:10.1016/j.nuclphysbps.2015.09.284.
\bibitem{lsnd} A. Aguilar et al. \textit{The LSND Collaboration} (2001) `Evidence for neutrino oscillations from the observation of $\APnue$ appearance in a $\APnum$ beam', Phys. Rev. D, \textbf{(64)} 112007. doi:10.1103/PhysRevD.64.112007.
\bibitem{miniboone} A. A. Aguilar-Arevalo et al. \textit{The MiniBooNE Collaboration} (2013) `Improved search for $\APnum \rightarrow \APnue$ Oscillations in the MiniBooNE experiment', Phys. Rev. Lett., 110\textbf{(16)}. doi:10.1103/physrevlett.110.161801.
\bibitem{reactoranomaly} G. Mention et al. (2011) `The Reactor Antineutrino Anomaly', Phys. Rev. D, \textbf{(83)} 073006. doi:10.1103/PhysRevD.83.073006.
\bibitem{gallex} Kaether et al. (2010) `Reanalysis of the Gallex solar neutrino flux and source experiments', Phys. Lett. B, 685\textbf{(1)}, 47-54. doi:10.1016/j.physletb.2010.01.030.
\bibitem{sage} J. N. Abdurashitov et al. \textit{The SAGE Collaboration} (2009) `Measurement of the solar neutrino capture rate with gallium metal. III. Results for the 2002–2007 data-taking period', Phys. Rev. C, \textbf{(80)} 015807. doi:10.1103/PhysRevC.80.015807.
\bibitem{gallium} C. Giunti \& M. Laveder (2011) `Statistical significance of the gallium anomaly', Phys. Rev. C, \textbf{(83)} 065504. doi:10.1103/PhysRevC.83.065504.

\bibitem{sciboone} G. Cheng et al. \textit{The MiniBooNE and SciBooNE Collaborations} (2012) `Dual baseline search for muon antineutrino disappearance at 0.1 eV$^2 < \Delta m^2 < $100 eV$^2$', Phys. Rev. D, \textbf{(86)} 052009. doi:10.1103/PhysRevD.86.052009.
\bibitem{karmen} B. Armbruster et al. \textit{The KARMEN Collaboration} (2002) `Upper limits for neutrino oscillations $\overline{\nu}_\mu \rightarrow \overline{\nu}_e$ from muon decay at rest', Phys. Rev. D, 65\textbf{(11)}. doi:10.1103/PhysRevD.65.112001.
\bibitem{dayabay} F. P. An et al. \textit{The Daya Bay Collaboration} (2016) `Improved Search for a Light Sterile Neutrino with the Full Configuration of the Daya Bay Experiment'. Phys. Rev. Lett., 15\textbf{(117)} 151803. doi:10.1103/PhysRevLett.117.151802.
\bibitem{icecube} M. G. Aartsen et al. \textit{The IceCube Collaboration} (2017) `Search for sterile neutrino mixing using three years of IceCube DeepCore data', arXiv:1702.05160.
\bibitem{minos} P. Adamson et al. \textit{The MINOS Collaboration} (2016) `Search for Sterile Neutrinos Mixing with Muon Neutrinos in MINOS', Phys. Rev. Lett., 15\textbf{(117)} 151803. doi:10.1103/PhysRevLett.117.151803.
\bibitem{planck} P. A. R. Ade et al. (2015) `Planck 2015 results. XIII. Cosmological parameters', Astron. Astrophys. \textbf{(594)} A13. doi:10.1051/0004-6361/201525830.
\bibitem{rh} M. Drewes (2013) `The phenomenology of right handed neutrinos', Int. Jour. of Mod. Phys. E, 22\textbf{(08)} 1330019. doi:10.1142/S0218301313300191.
\bibitem{steveyboyd} S. Boyd (2014) `Neutrino Mass and Direct Measurements' Available at: \url{https://www2.warwick.ac.uk/fac/sci/physics/staff/academic/boyd/stuff/neutrinolectures/lec_neutrinomass_writeup.pdf} [Accessed 24th April 2017].
\bibitem{seesaw} H. Zhang (2012) `Light sterile neutrino in the minimal extended seesaw', Phys. Lett Sect. B, 2-5\textbf{(714)}, 262-266. doi:10.1016/j.physletb.2012.06.074.
\bibitem{gut} J. C. Pati \& A. Salam (1974) `Lepton number as the fourth ``color"', Phys. Rev. D, 1\textbf{(10)}, 275. doi:10.1103/PhysRevD.10.275.
\bibitem{cosmology} M. Lucente (2016) `Implication of Sterile Fermions in Particle Physics and Cosmology', arXiv:1609.07081.
\bibitem{nucleosynthesis} G. Steigman (2012) `Neutrinos and big bang nucleosynthesis', Adv. in High Ener. Phys., \textbf{(2012)} 268321. doi:10.1155/2012/268321.
\bibitem{dm} S. Tremaine \& J.E. Gunn (1979) `Dynamical Role of Light Neutral Leptons in Cosmology', Phys. Rev. Lett., 1\textbf{(42)}, 407.
\bibitem{whitepaper} K. N. Abazajian et al. (2012) `Light Sterile Neutrinos: A White Paper', arXiv:1204.5379.
\bibitem{cdm} G. B. Gelmini, E. Osoba \& S. Palomares-Ruiz (2010) `Inert-Sterile Neutrino: Cold or Warm Dark Matter Candidate', Phys. Rev. D, \textbf{(81)} 063529. doi:10.1103/PhysRevD.81.063529.
\bibitem{reactorflux} T. A. Mueller et al. (2011) `Improved predictions of reactor antineutrino spectra', Phys. Rev. C, 83\textbf{(5)} 054615. doi:10.1103/PhysRevC.83.054615.
\bibitem{Giunti2011a} C. Giunti, M. Laveder (2011) `3+1 and 3+2 sterile neutrino fits', Phys. Rev. D, 7\textbf{(84)} 073008. doi:10.1103/PhysRevD.84.073008.
\bibitem{bbnn} G. Mangano (2011) `A robust upper limit on $N_\text{eff}$ from BBN, circa 2011', 701\textbf{(3)}, 296–299. doi:10.1016/j.physletb.2011.05.075.
\bibitem{bbn} G. Steigman (2012) `Neutrinos and Big Bang Nucleosynthesis', Adv. High Energy Phys., \textbf{(2012)} 268321. doi:10.1155/2012/268321.
\bibitem{nsi} E. Akhmedov \& T. Schwetz (2010) `MiniBooNE and LSND data: non-standard neutrino interactions in a (3+1) scheme versus (3+2) oscillations', High Energ. Phys., 115\textbf{(2010)}. doi:10.1007/JHEP10(2010)115.
\bibitem{Kopp2013} Kopp et al. (2013) `Sterile neutrino oscillations: the global picture', JHEP 1305 5\textbf{(2013)}. doi:10.1007/JHEP05(2013)050.
\bibitem{Collin2016} G. Collin et al. (2016) `Sterile neutrino fits to short baseline data', Nuc. Phys. B, \textbf{(908)}, 354-365. doi:10.1016/j.nuclphysb.2016.02.024.
\bibitem{2017} S. Gariazzo et al. (2017) `Updated Global 3+1 Analysis of Short-BaseLine Neutrino Oscillations', arXiv:1703.00860.
\bibitem{reactorspectra} A.C. Hayes and P. Vogel (2016) `Reactor Neutrino Spectra', Ann. Rev. of Nucl. and Part. Sci., \textbf{(66)}, 219-244. doi:10.1146/annurev-nucl-102115-04482.
\bibitem{lar1nd} C. Adams et al. (2013), `LAr1-ND: Testing Neutrino Anomalies with Multiple LArTPC Detectors at Fermilab', C13-07-29.2 Proceedings, arXiv:1309.7987.
\bibitem{lar1nd1} N. McConkey (2013) `The LAr1-ND Experiment', Journal of Physics: Conference Series, \textbf{(650)}. doi:10.1088/1742-6596/650/1/012007.
\bibitem{microboone} R. Acciarri (2017) `Design and construction of the MicroBooNE detector', Jour. of Instr., \textbf{(12)}. doi:10.1088/1748-0221/12/02/P02017.
\bibitem{icarus}  M. Antonello et al. (2011) `Underground operation of the ICARUS T600 LAr-TPC: first results', Jour. of Instr., \textbf{(6)} P07011. doi:10.1088/1748-0221/6/07/P07011.
\bibitem{proposal} M. Antonello et al. \textit{MicroBooNE and LAr1-ND and ICARUS-WA104 Collaborations} (2015) `A Proposal for a Three Detector Short-Baseline Neutrino Oscillation Program in the Fermilab Booster Neutrino Beam', arXiv:1503.01520.
\bibitem{scibooneminiboone} K. B. M. Mahn et al. \textit{MiniBooNE and SciBooNE Collaborations} (2012) `Dual baseline search for muon neutrino disappearance at $\SI{0.5}{\electronvolt\squared}\textless\Delta m^2\textless\SI{40}{\electronvolt\squared}$', Phys. Rev. D, 3\textbf{(85)} 032007. doi:10.1103/PhysRevD.85.032007.
\bibitem{bnbflux} A. A. Aguilar-Arevalo et al. \textit{The MiniBooNE Collaboration} (2009) `Neutrino flux prediction at MiniBooNE', Phys. Rev. D, \textbf{(79)} 072002. doi:10.1103/PhysRevD.79.072002.
\bibitem{lartpc} C. Rubbia (1977) `The liquid-argon time projection chamber : a new concept for neutrino detectors', CERN-EP-INT-77-8.
\bibitem{image} J. Spitz \textit{The ArgoNeut Collaboration} (2009) `ArgoNeuT and the Neutrino-Argon Charged Current Quasi-Elastic Cross Section', J. Phys. Conf. Ser., \textbf{(312)} 072017. doi:10.1088/1742-6596/312/7/072017.
\bibitem{mip} S. Gollapinni \textit{The MicroBooNE Collaboration} (2015) `Accelerator-based Short-baseline Neutrino Oscillation Experiments', CIPANP2015-GOLLAPINNI. arXiv:1510.04412.
\bibitem{mip2} A. Rubbia (2013) `Future liquid Argon detectors', Nucl.Phys.Proc.Suppl. 235-236\textbf{(2013)}, 190-197. doi:10.1016/j.nuclphysbps.2013.04.010
\bibitem{microbooneee} R. Acciarri et al. \textit{The MicroBooNE Collaboration} (2016) `Convolutional Neural Networks Applied to Neutrino Events in a Liquid Argon Time Projection Chamber', Jour. of Instr., 12\textbf{(3)} P03011. doi:10.1088/1748-0221/12/03/P03011.
\bibitem{argoneut} C. Anderson et al. (2012) `The ArgoNeuT Detector in the NuMI Low-Energy beam line at Fermilab', Jour. of Instr., \textbf{(7)} P10019. doi:10.1088/1748-0221/7/10/P10019.
\bibitem{argoneut1} R. Acciarri et al. \textit{The ArgoNeuT Collaboration} (2016) `First Observation of Low Energy Electron Neutrinos in a Liquid Argon Time Projection Chamber', Phys. Rev. D, FERMILAB-PUB-16-466-ND, arXiv:1610.04102.
\bibitem{protodune} F. Cavanna \textit{NP04 Collaboration} (2016) `Yearly report on ProtoDUNE Single Phase NP04', Tech. Rep., CERN-SPSC-2016-018,SPSC-SR-185. Available at: \url{https://cds.cern.ch/record/2144868}. [Accessed 3rd April 2017].
\bibitem{microbooneres} P. Hamilton (2016) `First Measurement of Neutrino Interactions in MicroBooNE', FERMILAB-CONF-16-548-ND, arXiv:1611.00820.
\bibitem{sbn2} D. Cianci et al. (2017) `Prospects of Light Sterile Neutrino Oscillation and CP Violation Searches at the Fermilab Short Baseline Neutrino Facility', arXiv:1702.01758.
\bibitem{giunti2013} C. Giunti et al. (2013) `Pragmatic View of Short-Baseline Neutrino Oscillations', Phys. Rev. D, \textbf{(88)} 073008, doi:10.1103/PhysRevD.88.073008.
\bibitem{rebel} B. Rebel (2009) `General form of the Unitary Mixing Matrix for 4 Neutrino Mass Eigenstates', MINOS internal papers. [Accessed 21st February 2017].
\bibitem{lbne} A. Heavey (2012) `LBNE Long-Baseline Neutrino Experiment Document 5235-v9', LBNE CDR (DocDB 5235). Available at: \url{http://lbne2-docdb.fnal.gov/cgi-bin/ShowDocument?docid=5235}. [Accessed 2nd April 2017].
\bibitem{microboonedata} \textit{The MicroBooNE Experiment} (2015) `Index of public plots and other data representations', Available at: \url{http://www-microboone.fnal.gov/public_plots/index.html}. [Accessed 29th March 2017].



\end{thebibliography}
\end{document}